\documentclass{elsart}
\usepackage{amsmath,epsfig}
\begin{document}
\begin{frontmatter}
\title{Perturbed soliton excitations in  DNA double helix }

\author{M.~Daniel\corauthref{cor1}$^{a,b}$},
\ead{daniel@cnld.bdu.ac.in} 
\author{V.~Vasumathi$^{a,b}$}

\corauth[cor1]{Corresponding Author.  Fax:+91-431-2407093}
\address{a. Centre for Nonlinear Dynamics, Department of Physics, 
Bharathidasan University, Tiruchirappalli - 620 024, India.\\
b. The Abdus Salam International Centre for Theoretical Physics,
Strada Costiera-11,
34014 Trieste, Italy}
\date{}

\begin{abstract}
   We study  nonlinear dynamics of  inhomogeneous DNA double helical chain under
   dynamic plane-base rotator model by considering angular rotation of bases 
   in a plane normal to the helical
   axis. The DNA dynamics  in this case is found to be governed
    by a perturbed sine-Gordon equation while taking into account the interstrand hydrogen
    bonding energy  between bases and the intrastrand inhomogeneous stacking energy and by making an analogy
     with the Heisenberg model of the Hamiltonian of an inhomogeneous anisotropic spin ladder with
     ferromagnetic legs and antiferromagnetic rung coupling. In the homogeneous limit the dynamics is
     governed by the kink-antikink soliton of the sine-Gordon equation which represents the
      formation of open
     state configuration in DNA double helix. The effect of inhomogeneity 
     in stacking energy in the form  of
    localized and periodic variations  on the  formation of open states  in DNA 
     is studied under perturbation.  The perturbed soliton is obtained using a multiple scale soliton perturbation theory by solving
    the associated linear eigen value problem and by constructing the complete set of eigen functions. 
    The inhomogeneity in stacking energy is found to modulate the width  and speed of the soliton 
     depending on the nature of inhomogeneity. Also it  introduces fluctuations 
      in the form of train of
     pulses or periodic oscillations   in the open state
    configuration. 
\end{abstract}

\end{frontmatter}

\section{Introduction}
   A number of theoretical models have been proposed to describe  nonlinear molecular
    excitations in DNA  double helix which plays an important role in the
 conservation and transformation of genetic information in  biological
 systems \cite{ref1}. 
    These theoretical models are based on
     longitudinal and transverse motions, as well as bending,
      stretching and rotations \cite{ref2,ref3}.  
      Among the different motions,
     the rotational motion of bases in DNA is found to contribute more towards the opening 
     of base pairs.           
     The first contribution towards nonlinear dynamics of DNA was made by Englander and his
     co-workers \cite{ref4} who studied the dynamics of DNA open states by taking into account only the rotational
     motion of nitrogenous bases, which made the main contribution towards the formation of open states.  Yomosa \cite{ref6,ref7} developing this idea further proposed a dynamic plane base rotator model which
     is a generalized version of the Frenkel-Kontrova \cite{ref8} model that was later  improved by
     Takeno and Homma \cite{ref9,ref10} in which attention was paid to the degree of freedom, characterizing base
     rotations in the plane perpendicular to the helical axis around the backbone structure.  In the above, the DNA dynamics was governed by the completely integrable sine-Gordon model admitting kink-type solitons. Then
     Peyrard and Bishop \cite{ref11}  studied the process of denaturation in which only the
     transverse motion of
     bases along the hydrogen bond was taken into account. There was one more
     model studied by   
     Christiansen and his colleagues (see for e.g.\cite{ref12}) using Toda lattice model  in which two types of internal motions namely, transverse motion
     along the hydrogen bond direction and longitudinal motion along the backbone direction were found 
     to contribute to DNA denaturation process in terms of travelling solitary waves and standing waves.  		
  These  localized nonlinear excitations  further explain  conformation transition 
  \cite{ref14,ref15,ref16},
      long range interaction of kink solitons in the double chain \cite{ref17,ref18}, regulation of transcription
       \cite{ref17,ref19},
      denaturation \cite{ref11} and
      charge transport in terms of polarons and bubbles \cite{ref21}. Some of them have been successfully used for interpreting
       experimental data related to
       microwave
     absorption \cite{ref22,ref23}. Further developments, in this approach for  several years
     was limited  to small
     improvements of the models  involving only numerical methods of simulation of the internal
     dynamics of DNA \cite{ref24,ref25,ref13}. Also, bubbles   \cite{ref20} discrete breathers \cite{ref26,ref27,ref28} and non-breathing compacton-like modes were obtained by
     solving the DNA-lattice model \cite{ref29}.  Eventhough, the rotation of  bases in DNA is mainly due to thermal forces the thermal fluctuation in DNA dynamics has been  introduced through random forces  in the recent past. For instance, Yakushevich et al \cite{ref13} has shown that topological solitons of the DNA chains are stable with respect to thermal oscillations. Since random thermal forces introduce only small fluctuations, it is not included in the present study. Thus, the study of nonlinear  excitations in DNA molecular chain has become an important
 task since it is related to its major functions.\\
 	
  In all  the above  studies, DNA double helix with homogeneous stiff strands has been considered for the
  analysis. However, in nature,
   the presence of different sites along the strands such as promotor, coding, terminator etc. each
   of which having a very specific sequence of bases and particular functions makes the strands
   site-dependent or inhomogeneous(soft). Also, defects  caused due to  external molecules  
     in the sequence and  the presence of
    abasic site-like nonpolar mimic of thymine lead to inhomogeneity \cite{ref30,ref31}. When included, the DNA dynamics is governed by an inhomogeneous perturbed sine-Gordon equation and thus the problem boils down to solving the same and finding  perturbed solitons. Also, in a different context, in the recent times, the study of wave propagation, especially solitons through inhomogeneous or disordered media assumed lot of interest \cite{ref5}. For instance, kink-impurity interaction and its scattering in the sine-Gordon model was studied in detail by Zhang Fei et al \cite{ref32}. With this in mind, in the present paper we study the nonlinear
   dynamics of  DNA double helix with inhomogeneous strands by considering a plane
     base rotator model along the lines of Yomosa. 
 	The paper is organized as follows.
	 In section 2 we introduce a dynamic plane base rotator model  
	for the site-dependent DNA double helix  and derive the nonlinear dynamical equation.
	 In section 3, we study the effect of inhomogeneity in stacking energy on the open state  of DNA in
	 terms of kink-antikink solitons by solving a perturbed sine-Gordon equation using a
	 multiple scale  soliton perturbation theory. The results are concluded in section 4.
	 Detailed evaluation of few integrals using residue theorem is given in Appendix.
 
\section{Plane-base rotator Model and Dynamical equation}     
    
      		 In Fig.1a  we have presented a schematic 
     structure of the B-form
      DNA double helix. Here $S$ and $S'$  represent the two complementary strands in the DNA double helix.
      Each arrow in the figure represents the direction of the base attached
     to the strand and the dots between arrows represent  the net hydrogen bonding effect between the 
     complement bases. While a horizontal
     projection of the $n^{th}$ base pair in the XY-plane is represented in Fig.1b, in Fig.1c, we have given the
     projection of the same in the XZ-plane. 
       The Z-axis is chosen along the helical axis of the DNA.
     In Fig.1b, $Q_n$ and  $Q'_n$ denote the  tip of the $n^{th}$ bases belonging to  the
     complementary strands $S$ and $S'$. $P_n$ and $P'_{n}$ represent the points where the bases in  the $n^{th}$ base 
     pair are attached to the strands $S$ and
     $S'$ respectively. Let $(\theta_n,\phi_n)$ and $(\theta'_n,\phi'_n)$ represent the
     angles of rotation of the bases in the  $n^{th}$ base pair around the points $P_n$ and $P'_n$ in the XZ and
     XY-planes respectively.\\
\begin{figure}      
\begin{center}
\epsfig{file=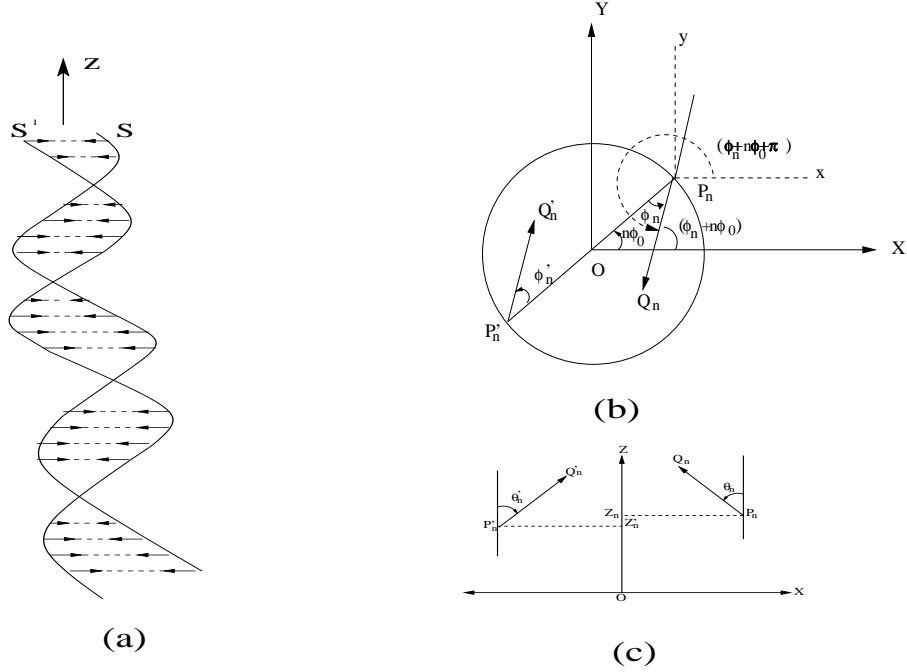,height =9cm, width=12cm}
\caption{(a) A schematic structure  B-form  DNA double helix. (b) A horizontal projection of the $n^{th}$ base
pair in the XY-plane. (c) A projection of the $n^{th}$ base pair in the XZ-plane.}
\end{center}
\end{figure}
       The conformation and stability of  DNA double helix is  mainly determined by the stacking
     energy between the intrastrand adjacent bases, the hydrogen bonding energy between the interstrand
     complementary bases and other energies. From a heuristic argument it was assumed that the interstrand 
     base-base interaction or
     hydrogen bonding energy of the given base  pair  depends on  the distance between them. Thus, from Fig.1b we
     can write  down the square of the distance between the edges of  the arrows $(Q_nQ'_n)^{2}$
      as
\begin{eqnarray}
 ( Q_nQ'_n)^{2}&=&2+4 r^{2}+(z_n-z'_n)^2+2 (z_n-z'_n)
\left(\cos\theta_n -\cos\theta'_n\right) \nonumber\\
&&-4r\left[\sin\theta_n\cos\phi_n
+\sin\theta'_n\cos\phi'_n\right]
+2\left[\sin\theta_n\sin\theta'_n
\right.\nonumber\\
&&\times\left(\cos\phi_n\cos\phi'_n+\sin\phi_n\sin\phi'_n\right)
\left.-\cos\theta_n\cos\theta'_n\right],\label{eq1}
\end{eqnarray}
where `$r$' is the radius of the circle depicted in Fig.1b.
 The base-base interaction energy can be understood in a
more clear and  transparent way by introducing  quasi-spin operators ${\bf
S_n}=(S_n^{x}, S_n^{y}, S_n^{z})$ and $ {\bf S'_n}=(S_n^{'x}, S_n^{'y}, S_n^{'z})$ in the form \\
\begin{subequations}
\begin{eqnarray}
{ S_n^{x}}  =  \sin\theta_n\cos\phi_n,\quad
{ S_n^{y}} =  \sin\theta_n\sin\phi_n,\quad
{ S_n^{z}}  =  \cos\theta_n,\\
{ S_n^{'x}}  =  \sin\theta'_n\cos\phi'_n,\quad
{ S_n^{'y}} =  \sin\theta'_n\sin\phi'_n,\quad
{ S_n^{'z}} =  \cos\theta'_n, 
\end{eqnarray}
\end{subequations}
for the $n^{th}$ bases in the $S^{th}$ and $S^{'th}$  strands respectively. In view of
this, Eq.~(\ref{eq1}) can 
be written in terms of the given spin operators as follows.
\begin{eqnarray}
(Q_n Q'_n)^2=2+4r^2+2\left[S_n^{x} S_n^{'x}+S_n^{y} S_n^{'y}
 -S_n^{z} S_n^{'z}\right]
-4r\left[S_n^{x}+S_n^{'x}\right].\label{eq3}
\end{eqnarray}
While writing Eq.(\ref{eq3}) we have neglected the longitudinal compression  along the direction of the
helical axis thereby choosing $z_n=z'_n$.
It may be noted that the form of $({Q_nQ'_n})^2$ given in Eq.(\ref{eq3}) is the same as the Hamiltonian
for  a generalized form of the Heisenberg spin model. Therefore, the intrastrand base-base interaction
 in DNA can be
written using the same consideration. It is known that stacking or base-base interaction is a dominant force that stabilizes the DNA double helix. It is much stronger than the hydrogen bonding force and in fact the stacking between adjacent bases often contributes more than half of the total free energy of the  base pairs \cite{ref32a}. Several non-covalent forces including dipole-dipole interaction, van der Waals force etc stabilize stacking in DNA. It is also reasonable to think that if such a quasi-spin model
 can be used in this problem, the double strand DNA and the rung-like base pairs can be
conceived as an anisotropic  coupled spin chain model or spin ladder. This is schematically
presented in Fig.2. In the figure,  $S$ and $S'$ which previously represented  two strands of the DNA 
here correspond to two ferromagnetic lattices of a spin ladder with
antiferromagnetic coupling among the rungs. In the case of spin chains each arrow represents 
the magnetic moment corresponding to a group of atoms
at that lattice point. Due to antiferromagnetic  type of rung coupling the arrows  in the
lattice  $S$ and $S'$ are marked anti-parallel to each other. Here Z-direction (i.e) the 
direction of the helical axis is chosen as the easy axis of
magnetization in the spin chain.  Further, the spin-spin exchange interaction is restricted.  to
 the nearest neighbours, (i.e) the $n^{th}$
spin is coupled to the spins at the $(n+1)^{th}$ and $(n-1)^{th}$ sites.
\begin{figure}
\begin{center}
\psfig{file=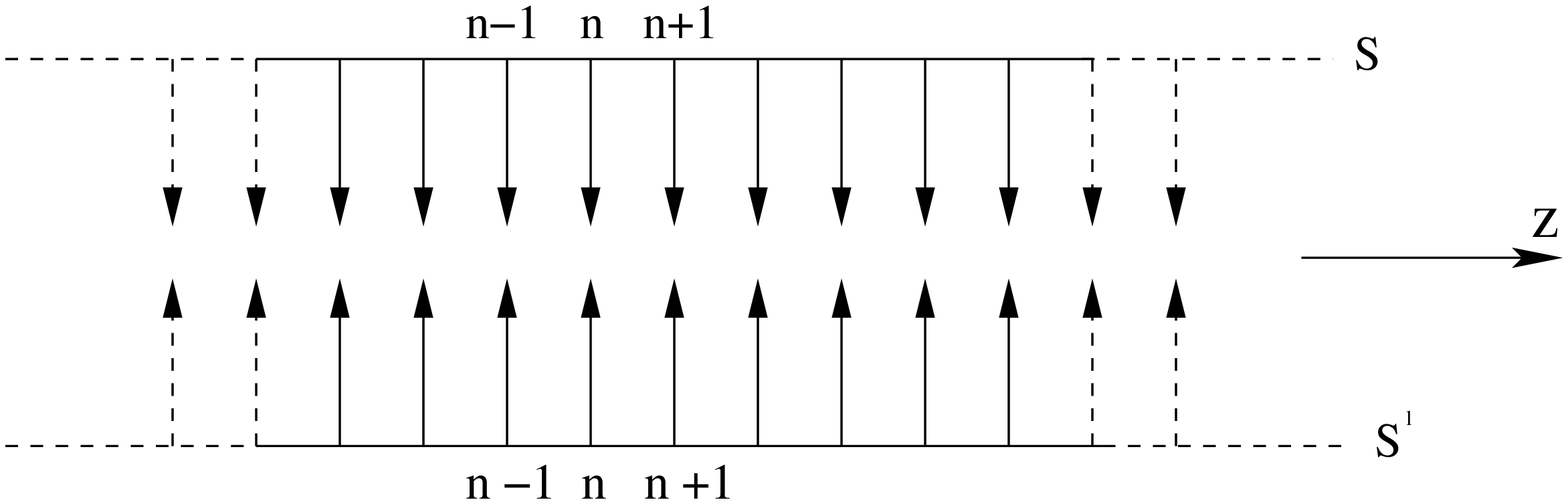,height =3cm, width=9cm}
\caption{A schematic representation of  DNA  as 
an anisotropic  coupled spin chain model or spin ladder.}
\end{center}
\end{figure} 
\\  

With this consideration we use the following  Heisenberg model of the
Hamiltonian for an anisotropic  coupled spin chain model or 
 spin ladder with site-dependent or inhomogeneous ferromagnetic-type
exchange interaction  between nearest neighbouring spins in the same lattice (equivalent to coupling among bases in 
the same strand i.e. intrastrand interaction) and antiferromagnetic rung-coupling (between
  bases belonging to the complementary strands i.e. interstrand interaction).
\begin{eqnarray}
H&=&\sum_n \left[-J f_n\left({S^{x}_{n}} {S^{x}_{n+1}}+{S^{y}_{n}}
{S^{y}_{n+1}}\right)-K f_n\right.
{S^{z}_{n}} {S^{z}_{n+1}}-J' f'_n\left({S^{'x} _{n}} {S^{'x}_{n+1}}\right.\nonumber\\
&&\left.+{S^{'y} _{n}} {S^{'y}_{n+1}}\right)
-K' f'_n {S^{'z}_{n}} {S^{'z}_{n+1}}
+\eta\left(S_n^{x}
{S_n^{'x}} +{S^{y}_n} {S^{'y}_n}\right)+\mu {S_n^{z}} {S^{'z}_n}\nonumber\\
&&
\left.+ A (S_n^{z})^{2}+A'
({S^{'z}_n})^{2}\right]. \label{eq4} 
\end{eqnarray}
In the above Hamiltonian  $J$ and $J^{'}$ correspond to the intrastrand interaction constant or the
 stacking energy between the $n^{th}$ base and its nearest neighbours in the plane normal to the helical 
 axis in the strands $S$ and $ S'$ respectively.
  When K and $K'$ are not equal to J and $J'$ respectively, they introduce anisotropy in the intrastrand
  interaction. $\mu$ and $\eta $  represent a measure of 
interstrand interaction or hydrogen bonding energy between the bases of similar sites in both the
strands  
 along the direction of the helical axis and in a plane
 normal to it respectively. Here we have assumed that there exist on an average  almost uniform
 interactions
 $A $ and $A^{'}$ that assume positive values which are the uniaxial anisotropy coefficients 
leading to rotation of the bases in a plane normal to the helical axis.  The quantities 
$ f_n$ and $f'_n$  in Hamiltonian ~(\ref{eq4}) indicate that the intrastrand stacking energy
between bases in the $S^{th}$ and $S^{'th}$ strand varies  in a specified site dependent 
 fashion which leads to  inhomogeneity in the DNA
 double helical 
chain. In general $f_n$ and $f'_n$ may take different values for different base sequences.
 The inhomogeneity in DNA double
helix may arise due to any one of the  reasons mentioned in the previous section.  The effect of
 inhomogeneity in DNA nonlinear dynamics has been studied  in the past by several
    authors in various contexts such as  variational mass density of bases
    \cite{ref33} and defect \cite{ref26}. \\
    
 To proceed further we rewrite  Hamiltonian~(\ref{eq4})
in terms of the variables $(\theta_n,\phi_n)$ and $ (\theta'_n,\phi'_n)$ using
 Eqs.(2) and obtain
\begin{eqnarray}
H&=&\sum_n\left[-J f_n \sin\theta_n\sin\theta_{n+1} \cos
(\phi_{n+1}-\phi_n)\right.
 -K f_n \cos\theta_n\cos\theta_{n+1}\nonumber\\
&&-J' f'_n\sin\theta'_n\sin\theta'_{n+1}  
\cos (\phi'_{n+1}-\phi'_n)-K' f'_n \cos\theta'_n\cos\theta'_{n+1}\nonumber\\
&&+\eta \sin\theta_n\sin\theta'_n 
\cos (\phi_n-\phi'_n)+\mu\cos\theta_n\cos\theta'_n+A\cos\theta_n^{2}\nonumber\\
&&\left.+A'
{\cos\theta'_n}^{2}\right].\label{eq5}
\end{eqnarray}
The quasi-spin model thus  introduced implies  that the dynamics of  bases in DNA can be described by the
following equations of motion.
\begin{eqnarray}
\dot\theta_n=\frac{1}{\sin\theta_n} \frac{\partial{H} } {\partial {\phi_n}},~
\dot\phi_n=\frac{-1}{\sin\theta_n}\frac{\partial{H}} {\partial {\theta_n}} ,~ 
\dot\theta'_n=\frac{1}{\sin\theta'_n} \frac{\partial{H} } {\partial  {\phi'_n}},~
\dot\phi'_n=\frac{-1}{\sin\theta'_n}\frac{\partial{H}} {\partial {\theta'_n}} .\label{eq6}
\end{eqnarray}
In Eq.(\ref{eq6})  the overdot represents  time derivative.
When the anisotropy energies $A$ and $A'$ are much larger than the other 
interactions, (i.e) when $A,A'>>J,J',K,K',\eta,\mu$,  then the equations of
motion ~(\ref{eq6}) on substituting the Hamiltonian ~(\ref{eq5}) become
\begin{eqnarray}
\dot \phi_n= 2 A \cos\theta_n,~ \dot\phi'_n= 2 A'\cos\theta'_n. \label{eq7}
\end{eqnarray}
The other two equations in (\ref{eq6}) satisfy identically.
 Using Eq.~(\ref{eq7}) in the Hamiltonian ~(\ref{eq5}) we obtain
\begin{eqnarray}
H&=&\sum_{n}\left[\frac{I}{2} {\dot\phi_n}^{2}+\frac{I'}{2}{\dot\phi_n}^{'2}
-J f_n \sin\theta_n\sin\theta_{n+1}\right.
 \cos(\phi_{n+1}-\phi_n)
 -K f_n \cos\theta_n\nonumber\\
&&\times\cos\theta_{n+1}-J' f'_n\sin\theta'_n\sin\theta'_{n+1}  
\cos (\phi'_{n+1}-\phi'_n)
-K' f'_n \cos\theta'_n\cos\theta'_{n+1}\nonumber\\
&&+\eta \sin\theta_n\sin\theta'_n 
\left.\cos (\phi_n-\phi'_n)+\mu\cos\theta_n\cos\theta'_n\right],\label{eq8}
\end{eqnarray} 
 where $I=\frac{1}{2A}$ and $ I'=\frac{1}{2A'}$.
 The  above Hamiltonian  can be rewritten  in the limits of plane-base rotator
 $(\theta_n=\theta'_n=\pi/2)$  and  absolute minima of
 potential as
 \begin{eqnarray}
 H&=&\sum_n\left[ \frac{I}{2} {\dot\phi_n}^{2}+\frac{I'}{2} {\dot\phi_n}^{'2} +J f_n
 \left[1-\cos (\phi_{n+1}-\phi_n)\right]\right.\nonumber\\
 &&+J' f'_n \left[1-\cos  (\phi'_{n+1}-\phi'_n)\right]
\left. -\eta\left[1-\cos  (\phi_n-\phi'_n)\right]\right]. \label{eq9}
 \end{eqnarray}
 It may be noted that the above limit corresponds to the XY-spin model of two coupled  inhomogeneous
 ferromagnetic spin system.
  In  Hamiltonian (\ref{eq9}) the first two terms represent the kinetic energies of the rotational motion
  of the $n^{th}$  nucleotide bases accompanied by the potential energy associated with the $n^{th}$ nucleotide sugar
  and phosphate and its complementary unit around the axes at $P_n$ and $P'_n$ (see Fig.1b).
    $I$ and $I'$ are the moments of inertia  of the
  nucleotides around the axes at $P_n$ and $P'_n$ respectively. It may be further noted that  in the new Hamiltonian the term proportional
  to $\mu$ vanishes.
  Now, using  Hamiltonian (\ref{eq9}), the  Hamilton's equations of motion can be  immediately written as
  \begin{subequations}
  \begin{eqnarray}
  I\ddot\phi_n&=&J \left[f_n \sin (\phi_{n+1}-\phi_n)-f_{n-1}\sin
  (\phi_n-\phi_{n-1})\right]
  +\eta \sin (\phi_n-\phi'_n), \label{eq10a}\\
  I'\ddot\phi'_n&=&J' \left[f'_n \sin (\phi'_{n+1}-\phi'_n)-f'_{n-1}\sin
  (\phi'_n-\phi'_{n-1})\right]
  +\eta \sin (\phi'_n-\phi_n).\label{eq10b}
  \end{eqnarray}  
  \end{subequations}
  Eqs.(\ref{eq10a}) and (\ref{eq10b}) describe the dynamics of DNA in a plane-base rotator model at the discrete level
   while  considering the dominant angular rotation of the bases  in a plane 
   normal to the helical axis and ignoring all other small motions of the bases.\\
  
     It is expected that in the B-form of DNA double helix   the difference  in the  angular rotation of  bases with respect to 
     neighbouring bases along the two strands is  small \cite{ref9,ref10}. Hence we assume that 
$\sin(\phi_{n\pm 1}-\phi_n)\approx(\phi_{n\pm 1}-\phi_n)$ and
$  \sin(\phi'_{n\pm 1}-\phi'_n)\approx(\phi'_{n\pm 1}-\phi'_n)$
 in Eqs.(\ref{eq10a},~b). Also, as the length of the DNA chain is very large  involving several thousands of base pairs compared to the distance between the
 neighbouring bases along the strands we  make a continuum approximation  as done by several authors in the past \cite{ref2,ref3,ref4,ref6,ref7,ref8,ref9,ref10,ref11} which is valid in the long wavelength, low temperature limit 
  by  introducing two fields of rotational angles,
 $\phi_n(t)\rightarrow\phi(z,t)$, $\phi'_n(t)\rightarrow\phi'(z,t)$ and 
 two inhomogeneous stacking fields, $f_{n}\rightarrow f(z)$  and  $f'_{n}\rightarrow f'(z)$  along with 
 the following expansions.
 \begin{eqnarray}
 \phi_{n\pm1}=\phi(z,t)\pm a\frac{ \partial{\phi}}{\partial{z}}+\frac{
 a^{2}}{2!}\frac{\partial^{2}{\phi}}{{\partial{z}}^{2}}\pm...,
 f_{n\pm 1}=f(z)\pm a \frac{\partial{f}}{\partial{z}}+\frac{ a^{2}}{2!}
 \frac{\partial^{2}{f}}{{\partial{z}}^{2}}\pm...,\label{eq 11}
 \end{eqnarray} 
where `$a$' is the lattice parameter along both
  the strands $S$ and $S'$. In a similar way we write down expansions for $\phi'_{n\pm 1}$ and $f'_{n\pm 1}$.
  As the inhomogeneity here is site-dependent and associated with the bases themselves, we have chosen same
  lattice parameter `$a$' for both $\phi_{n\pm 1}$ and $f_{n\pm 1}$.
   Under this  continuum approximation  the equations of motion (\ref{eq10a},~b) upto O($a^{2}$) is written as 
\begin{subequations}
\begin{eqnarray}
I\phi_{tt}&=& Ja^{2} \left[ f(z) \phi_{zz}+  f_z \phi_z\right]+ \eta\sin
(\phi-\phi'),\\
I' \phi'_{tt}&=& J'a^{2} \left[  f'(z) \phi'_{zz}+  f'_z \phi'_z\right] +\eta\sin
(\phi'-\phi).
\end{eqnarray}
\end{subequations}
In Eqs.~(12) the suffices $t$ and $z$  represent partial derivatives with respect 
to time $t$
and the spatial variable $z$ respectively.\\
 
In a DNA chain, the two strands are expected to exhibit similar type of macroscopic physical 
behaviour and
hence we assume that  $I=I',~ J=J'$ and $ f=f'$.
 In view of this,  Eqs.~(12)
 after rescaling  the time variable  as  
$\hat{t}=\sqrt{\frac{Ja^{2}}{I}} t
$ and choosing $\eta=-\frac{Ja^{2}}{2}$  for future convenience can be written as
\begin{subequations}
\begin{eqnarray}
\phi_{\hat t\hat t}&=&\left[f(z)\phi_{zz}+f_z \phi_{z} \right]
-\frac{1}{2} \sin (\phi-\phi'),\label{eq13a}\\
\phi'_{\hat t\hat t}&=&\left[f(z)\phi'_{zz}+f_z
\phi'_{z}\right]-\frac{1}{2} \sin (\phi'-\phi).\label{eq13b}
\end{eqnarray} 
\end{subequations}
It is more convenient to describe the transverse motion of  bases in  DNA strands in terms of the centre of
   mass co-ordinates. For this, we rewrite Eqs.~(\ref{eq13a}) and (\ref{eq13b}) 
   by subtracting and adding them respectively.
\begin{subequations}
\begin{eqnarray}
 (\phi-\phi')_{\hat t\hat t}&=&f(z)(\phi-\phi')_{zz}+f_z(\phi-\phi')_{z}
-\sin (\phi-\phi'),
\label{eq14a}\\
(\phi+\phi')_{\hat t\hat t}&=&f(z)(\phi+\phi')_{zz}+f_z(\phi+\phi')_{z}.  
\label{eq14b}
\end{eqnarray}
\end{subequations}
In a different context while studying the magnetoelastic effect induced by interaction between two
ferromagnetically coupled XY-spin chains in the static limit Dandoloff and 
Saxena \cite{ref36} 
obtained similar equations. To commence the open state configuration in DNA, the two complementary bases are expected to
 rotate in opposite  directions so
   that $\phi=-\phi'$, and in this case Eq.~(\ref{eq14b}) satisfies identically and
   Eq.~(\ref{eq14a}) becomes
\begin{eqnarray}
\Psi_{\hat t\hat t}=f \Psi_{zz}+f_z\Psi_z-\sin\Psi , \label{eq15}
\end{eqnarray}
   where $\Psi=2\phi$. Assuming small inhomogeneity along the strands by
   choosing $f(z)=1+\epsilon g(z)$ where $\epsilon$ is a small parameter
   and $g(z)$ a measure of the inhomogeneity, 
     Eq.~(\ref{eq15}) can be written as
\begin{eqnarray}
\Psi_{\hat t\hat t}-\Psi_{zz}+\sin\Psi=\epsilon \left[g(z)\Psi_{z}\right]_{z}.\label{eq16}
\end{eqnarray}
    Eq.~(\ref{eq16}) describes the dynamics of bases under rotation in a plane-base    rotator model of an inhomogeneous DNA double helical
    chain. When $\epsilon=0 $, 
     Eq.~(\ref{eq16}) reduces to the completely
     integrable sine-Gordon equation which admits kink and antikink-type of soliton solutions  and     
     hence we call Eq.(\ref{eq16})  as a perturbed sine-Gordon equation.
      The integrable sine-Gordon equation $(\epsilon=0$ case) was originally solved for N-soliton
      solutions using the most celebrated Inverse Scattering Transform (IST) method by Ablowitz and his
      co-workers \cite{ref37}. The kink and antikink  soliton solutions of the integrable sine-Gordon equation
(Eq.(\ref{eq16}) when $\epsilon=0$) are depicted in Figs.3a and 3b.
       The kink-antikink solitons of the sine-Gordon equation describe an open
       state  configuration in the DNA double helix. The formation of open state configuration in terms of
       kink-antikink pair in DNA double helical chain is schematically represented in Fig.3c. In this figure the base
        pairs are found to open locally in the form of kink-antikink 
	structure in each strand and propagate along the direction of the helical
	 axis.
\begin{figure}
\begin{center}
\epsfig{file=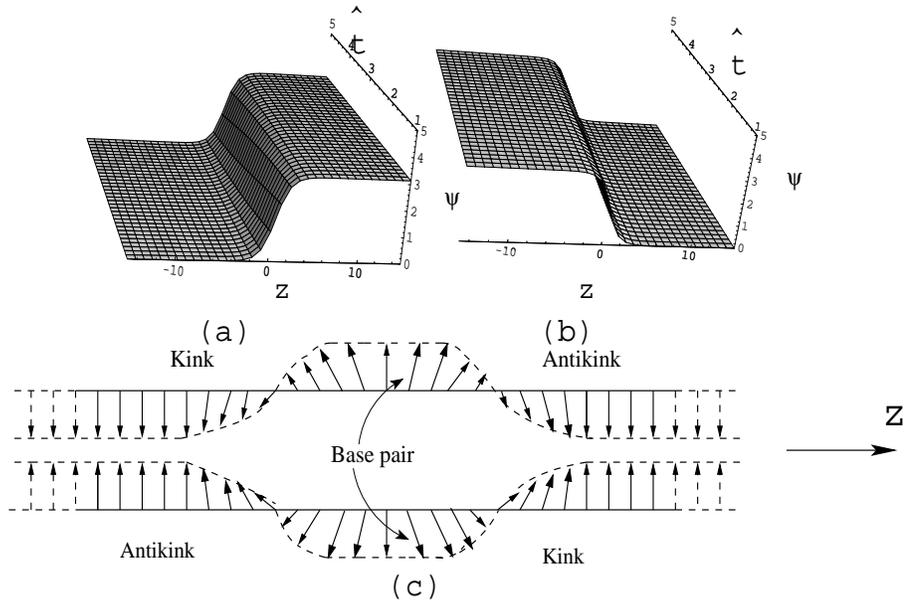,height =10cm, width=12cm}
\caption{(a) Kink  and (b) antikink soliton solutions of the sine-Gordon equation (Eq. (\ref{eq16}) when
$\epsilon=0$ ).
 (c) A sketch of the formation of open state configuration in terms of
kink-antikink solitons in DNA double helix.}
\end{center}
\end{figure}
 
\section{ Effect of  stacking energy Inhomogeneity on the Open State }  
\subsection{ A Perturbation approach}   
When  inhomogeneity in stacking is present  (i.e) when terms proportional to 
 $\epsilon$ are present in  Eq.(\ref{eq16}), the
	 inhomogeneity is expected to perturb the kink and antikink solitons corresponding 
	 to the open state of DNA. One of the most powerful techniques in dealing with perturbed soliton
	equations is the soliton perturbation theory  which is based on the IST method
	\cite{ref38,ref39}.
	 However, as the method is
	very sophisticated it is very difficult to use the same in
	several cases. In view of this, a direct method to study the soliton perturbation  was first
	introduced by Gorshkov and Ostrovskii \cite{ref40} and later   many  authors
	used different types of direct  methods to study  soliton perturbation (see for e.g. refs.
	\cite{ref41,ref42,ref43,ref44,ref45,ref46}).
	The  characteristic feature  of this method is that the  perturbed
	nonlinear equation is linearized by expanding  its solution about the
	unperturbed solution and  the eigen  functions for the
	operator  associated with the linearized equation are found out. The complete  solution is then written 
	in terms of these eigen functions \cite{ref47,ref48}. In 
	these methods the basic fact that the presence of perturbation not
	only modifies the shape of the soliton by a correction of  linear dispersion
	but also undergoes a slow time change of the soliton parameters have been acknowledged.
	 In this paper, we use one such direct perturbation method which is also dealt in reference 
	 \cite{ref49} in a different context
	   to solve the perturbed sine-Gordon equation (\ref{eq16}) and to understand
	    the effect of inhomogeneity in stacking
	 energy on the open state  configuration of DNA. 
	 The procedure we adapt here is based on the derivative
	expansion method to linearize the perturbed sine-Gordon equation in the
	co-ordinate frame attached to the moving frame. The parameters of the
	kink-antikink soliton are assumed to depend on a slow time scale in order to
	eliminate the secular terms. The linearized equations will be solved using
	the method of separation of variables which ultimately will be related to an eigen
	value problem, the eigen functions of which form the bases of the
	perturbed solution. The eigen functions contain information about
	the time dependence of the soliton parameters and help to calculate the
	perturbed soliton. In  the following we  use the above approach to find
	the perturbed  soliton solution of Eq.(\ref{eq16}). 
\subsection{Linearization of  the perturbed sine-Gordon equation}
     When the perturbation is absent (i.e) when $\epsilon =0$ in  	
     Eq.~(\ref{eq16}) the unperturbed integrable sine-Gordon equation 
      provides N-soliton solutions and the one soliton solution (see also Figs.3a and 3b is written as 
\begin{eqnarray} 
\Psi(z,\hat t) =4 arc\tan{exp[\pm m(z-v\hat t)]} ,\quad m=\frac{1}{\sqrt{1-v^2}}. \label{eq17}
\end{eqnarray}
In Eq.(\ref{eq17}) while the upper sign corresponds to kink soliton, the lower sign represents the antikink
soliton.
Here $ v$ and $m^{-1}$ are real parameters that determine the  velocity and width of the soliton 
respectively. In order to study the effect of perturbation the time variable $\hat t$ is transformed
into several variables as $t_n=\epsilon^{n} \hat t$ where n=0,1,2,... and $\epsilon$ is a very small parameter.
 In view of this, the
time derivative in Eq.~(\ref{eq16}) is replaced by the expansion
\begin{eqnarray}
\frac{\partial }{\partial \hat t}=\frac{\partial }{\partial t_0}+\epsilon~
\frac{\partial }{\partial t_1}+\epsilon^2 \frac{\partial }{\partial
t_2}+....\label{eq18}
\end{eqnarray}
Simultaneously $\Psi$ is expanded in an asymptotic series as
\begin{eqnarray}
\Psi=\Psi^{(0)}+\epsilon \Psi^{(1)}+\epsilon^2 \Psi^{(2)}+....\label{eq19}
\end{eqnarray}
Using the above expansions for $\hat{t}$ and $\Psi$ in Eq.~(\ref{eq16}) and equating
the coefficients of different powers of $\epsilon$, we obtain the following
equations.
\begin{subequations}
\begin{eqnarray}
\epsilon^{(0)}:\qquad \Psi^{(0)}_{t_0t_0}-\Psi^{(0)}_{zz}+\sin\Psi^{(0)}=0,
\qquad\qquad\qquad\qquad \qquad\quad\label 
{eq20a}\\
\epsilon^{(1)}:\qquad
\Psi^{(1)}_{t_0t_0}-\Psi^{(1)}_{zz}+\cos\Psi^{(0)}\Psi^{(1)}=g
\Psi^{(0)}_{zz}
+g_z \Psi^{(0)}_z-2 \Psi^{(0)}_{t_0 t_1},  \label {eq20b}
\end{eqnarray}
\end{subequations}
\quad etc.\\
The initial conditions for perturbation in the case of  the single soliton given in
Eq.~(\ref{eq17}) is written as  
\begin{subequations}
\begin{eqnarray}
\Psi^{(0)} (z,0)=4arc\tan\exp{mz},
\Psi^{(n)} (z,0)=\Psi^{(n)}_{t_{0}}(z,0)=0,~ n=0,1,....\quad\label {eq21b}
\end{eqnarray}
\end{subequations}
The equation obtained at order of $\epsilon^{(0)}$, (i.e) Eq.(\ref{eq20a}) is  just the  integrable sine-Gordon
equation for $\psi^{(0)}$, the single soliton solution of which can be written  from Eq.~(\ref{eq17})
immediately as
\begin{eqnarray}
 \Psi^{(0)}(z,t_0)= 4 arc\tan\exp\zeta,~\zeta= \pm m_{0}(z-\xi),~\xi_{t_0}=
 v_{0},\label{eq22}
\end{eqnarray}
where $v_{0}$ is the velocity of soliton in the $t_{0}$-time scale.
Due to perturbation, the soliton parameters  namely $m$ and $\xi$ are now treated
as functions of the slow time variables $t_0,t_1,t_2, .... $ However m is treated independent
of $t_0$. In view of the above,  Eq.~(\ref{eq20b}) becomes
\begin{subequations}
\begin{eqnarray}
 {\Psi^{(1)}_{t_0t_0}}-{\Psi_{zz}^{(1)}}+(1-2{sech^{2}} \zeta)\Psi^{(1)}=F^{(1)}
 (z) ,\label{eq23a}
\end{eqnarray}
where
\begin{eqnarray}
  F^{(1)} (z)= 2\left[g(z) sechz \right]_{z} + 4v_{0} sechz
\left[m_{t_{1}}+(m^{2}\xi_{t_{1}}-zm_{t_{1}})\tanh z
 \right].\quad \label{eq23b}
\end{eqnarray}
\label{eq23}
\end{subequations}
 While writing the above equation we have used  the result $\cos\Psi^{(0)}=1-2sech^{2}
 \zeta$  obtained from the solution of the unperturbed equation 
 (\ref{eq20a}).
In order to represent everything in a co-ordinate system moving with the soliton, we
transform $z=\frac{\zeta}{m}+v t_{0}$ and $t_0=t_0+\zeta$, so that Eqs.~(\ref{eq23}) 
 and the initial conditions given in Eq.
(\ref{eq21b}) can be written as
\begin{subequations}
\begin{eqnarray}
 {\Psi^{(1)}_{t_0t_0}}-2mv_{0}{\Psi_{t_0
 \zeta}^{(1)}}-{\Psi_{\zeta\zeta}^{(1)}}
 +(1-2{sech^{2}}\zeta)\Psi^{(1)}=F^{(1)}
 (\zeta,t_{0}), \label{eq24a}
\end{eqnarray}
 where
\begin{eqnarray}
  F^{(1)}(\zeta,t_{0})=2\left[g(\zeta) sech\zeta\right]_{\zeta}+4v_{0} sech\zeta
\left[m_{t_{1}}+(m^{2}\xi_{t_{1}}-\zeta
m_{t_{1}})\tanh\zeta
 \right],\label{eq24b}
\end{eqnarray}
 and
\begin{eqnarray}
 \Psi^{(1)}(\zeta,0)=0, \quad \Psi_{t_{0}}^{(1)} (\zeta,0)=0. \label{eq24c}
\end{eqnarray}
\end{subequations}
  We now introduce one more  transformation
 $\tau=\frac{t_0}{2m}-\frac{(1+v_{0})\zeta}{2}$ on the independent variable to
 eliminate the first term in the left hand side of Eq.~(\ref{eq24a}). Thus on
 using the above transformation  Eq.~(\ref{eq24a}) becomes 
 \begin{eqnarray}
 \Psi_{\tau\zeta}^{(1)}-{\Psi_{\zeta\zeta}^{(1)}}+(1-{sech^{2}}\zeta)\Psi^{(1)}=F^{(1)}
 (\zeta,\tau),\label{eq25}
  \end{eqnarray} 
  where $ F^{(1)}(\zeta,\tau)$ equals exactly the right hand side of Eq. (\ref{eq24b}).
 The solution of Eq.~(\ref{eq25}) is searched by assuming 
\begin{eqnarray}
\Psi^{(1)} (\zeta,\tau)=X(\zeta)T(\tau),\quad
F^{(1)} (\zeta,\tau)= X_{\zeta}(\zeta)H(\tau).\label{eq26}
\end{eqnarray}
Substituting Eq.~(\ref{eq26}) in Eq.~(\ref{eq25}) we obtain
\begin{eqnarray}
\frac{1}{X_{\zeta}}\left[X_{\zeta\zeta}+(2
sech^{2}\zeta-1)X\right]=\frac{1}{T}\left[T_{\tau}-H(\tau)\right].\label{eq27}
\end{eqnarray}
 In Eq.~(\ref{eq27}) the left hand side is independent of $\tau$  and the right hand side
 is independent of the variable $\zeta$. Hence we can equate the left and right hand
  sides  of Eq.~(\ref{eq27}) to a constant say $\lambda_{0}$ and write 
\begin{subequations}
\begin{eqnarray}
X_{\zeta\zeta}+(2 sech^{2}\zeta-1)X=\lambda_{0} X_{\zeta}, \label{eq28a}\\
T_{\tau}-\lambda_{0} T= H(\tau).\label{eq28b}
\end{eqnarray}
\label{eq28}
\end{subequations}
 Thus, the problem of constructing the
perturbed soliton at this moment turns out to be solving 
 Eqs.~(\ref{eq28a}) and (\ref{eq28b}) by constructing the eigen functions and 
finding the eigen values. It may
be noted that Eq.~(\ref{eq28a}) is a generalized eigen value problem which is
 not a self-adjoint eigen value problem and differing from the normal
eigen value problem with $X_{\zeta}$ in the right hand side instead of $X$ and
Eq.~(\ref{eq28b}) is a first order linear inhomogeneous differential
 equation which can be solved using known procedure.
\subsection{Solving the eigen value problem }
Before  actually solving the eigen value equation (\ref{eq28a}) 
we  first consider it
 in a more general form given by
\begin{eqnarray}
\qquad L_{1} X=\lambda \tilde{X} ,\qquad L_{1}=\partial_{\zeta\zeta}+ 2
sech^{2}\zeta -1,  \label{eq29}
\end{eqnarray}
 where $\lambda$ is the eigen value.
In order to find the adjoint eigen function to $X$, we consider another eigen value problem
\begin{eqnarray}
\qquad L_{2} \tilde{X}=\lambda X, \label{eq30}
\end{eqnarray}
where  the operator $L_{2}$ is still to be determined.
 The eigen  value equations (\ref{eq29}) and (\ref{eq30}) can be  combined to give
\begin{eqnarray} 
L_{2}L_{1} X=\lambda^{2} X, \quad
L_{1}L_{2} \tilde{X}=\lambda^{2} \tilde{X} .\label{eq31b}
\end{eqnarray}
  Since we already  know the form of $L_1$  as given in Eq.(\ref{eq29}), if
 we choose the operator $L_{2}$ as $ L_{2}=\partial_{\zeta\zeta}+ 6sech^{2}\zeta -1  $,
  it can  be verified that $L_{1}L_{2}$ is the adjoint of $L_{2}L_{1}$. Thus X and
  $\tilde{X}$ are expected to be adjoint  eigen functions.\\
  
 Now for solving the eigen value equations ~(\ref{eq29}) and (\ref{eq30}),  we
   choose the eigen functions $X$ and $\tilde{X}$ to be in the form
\begin{eqnarray}
X(\zeta,k)=p(\zeta,k) e^{ik\zeta},\quad
\tilde {X}(\zeta,k)=q(\zeta,k) e^{ik\zeta}, \label{eq32}
\end{eqnarray}
where $p(\zeta,k)$ and $q(\zeta,k)$ are assumed to have the asymptotic behaviour
 $p(\zeta,k)\rightarrow$ a constant  and $q(\zeta,k)\rightarrow$ a constant  as
  $\zeta\rightarrow \pm\infty$ and $k$ is the propagation constant. On using
 these asymptotic forms for $p(\zeta,k)$ and $q(\zeta,k)$ in 
 Eq.~(\ref{eq32})   and then substituting the  resultant
 $X(\zeta,k)$ and $\tilde {X}(\zeta,k)$ 
 in Eqs.~(\ref{eq29}) and (\ref{eq30}) we obtain the
 eigen value as 
 \begin{eqnarray}
\lambda =-(k^{2}+1). \label{eq33}
\end{eqnarray}
  On substituting  the exact forms of $X$ and $\tilde{X}$ from Eq.~(\ref{eq32}) in Eqs.~(\ref{eq29}) and (\ref{eq30}) we get the
  following set of ordinary differential equations for $p(\zeta,k)$ and
  $q(\zeta,k)$ .
\begin{subequations} 
\begin{eqnarray}
(L_{1}-k^{2})p+2ikp_{\zeta}+(1+k^{2})q=0, \label{eq34a}\\
(L_{2}-k^{2})q+ 2ikq_{\zeta}+(1+k^{2})p=0. \label{eq34b}  
\end{eqnarray}
\end{subequations}
For solving the above equations we expand $p(\zeta,k)$ and $q(\zeta,k)$ in 
  the following  series  \cite{ref49}.
\begin{subequations}
\begin{eqnarray}
p (\zeta,k)&=&p_0+p_1 \frac{\sinh\zeta}{\cosh\zeta}+p_2 \frac{1}{\cosh^{2}\zeta}+p_3
\frac{\sinh\zeta}{\cosh^{3}\zeta} 
+p_4\frac{1}{\cosh^{4}\zeta}+...,\label{eq35a}\\
q (\zeta,k)&=&q_0+q_1\frac{\sinh\zeta}{\cosh\zeta}+q_2 \frac{1}{\cosh^{2}\zeta}+q_3
\frac{\sinh\zeta}{\cosh^{3}\zeta} 
+q_4\frac{1}{\cosh^{4}\zeta}...,\qquad\label{eq35b}
\end{eqnarray}
\end{subequations}
where  the coefficients $p_j$ and $q_{j}$, j=0,1,2,... are functions of $k$ which are to be
determined. We  substitute the series expansions given in
Eqs.~(\ref{eq35a}) and (\ref{eq35b})
in Eqs.(34)  and collect the coefficients of $ 1,
\frac{\sinh\zeta}{\cosh\zeta}, \frac{1}{\cosh^{2}\zeta}$,... and obtain  the following algebraic equations.
\begin{subequations}
\begin{eqnarray}
 p_{0}=q_{0}, \quad p_{1}=q_{1}, \qquad\qquad\qquad \label{eq36a}\\
 2p_{0}+2ik p_{1}+(3-k^{2})p_{2}-4ik p_{3}+(1+k^{2})q_{2}=0,\label{eq36b}\\ 
6q_{0}+2ikq_{1}+(3-k^{2})q_{2}-4ik q_{3}+(1+k^{2})p_{2}=0,\label{eq36c}\\ 
-4ik p_{2}+(3-k^{2})p_{3}+(1+k^{2})q_{3}=0,\label{eq36d}\\
4q_{1}-4ik q_{2}+(3-k^{2})q_{3}+(1+k^{2})p_{3} =0.\label{eq36e}\\
etc.\qquad\qquad\qquad\qquad\qquad\qquad\qquad\qquad\nonumber
\end{eqnarray}
\end{subequations} 
 By assuming $p_{j}=q_{j}=0$ for $j\geq 3$,  and substituting these values in
 Eqs.(\ref{eq36d}) and (\ref{eq36e}), we obtain
 \begin{eqnarray}
 p_{2}=0,\quad q_{2}=-i\frac{q_{1}}{k}. \label{eq37}
 \end{eqnarray} 
 On substituting the results given in Eq.(\ref{eq37}) in Eqs. 
 (\ref{eq36b}) and (\ref{eq36c})
 we get
 \begin{eqnarray}
 p_{1}=q_{1}=-\frac{2ik p_{0}}{(1-k^{2})}, \quad q_{2}=-\frac{2p_{0}}{(1-k^{2})}.\label{eq38}
 \end{eqnarray}
 For our convenience we choose $p_{0}=q_{0}=c (1-k^{2})$ and on substituting this
 in the above equations, we obtain the other coefficients as follows.
\begin{eqnarray} 
p_{1}=q_{1}=-2ik,~p_{2}=0,~q_{2}=-2c.\label{eq39a} 
\end{eqnarray}
Here `$c$' is an arbitrary constant which will be determined. Using the above  values in Eqs.~(\ref{eq35a})
and (\ref{eq35b}), and then in Eq.~(\ref{eq32}),    we finally obtain 
\begin{subequations} 
\begin{eqnarray}
X(\zeta,k)&=&c (1-k^{2}-2ik\tanh\zeta) e^{ik\zeta}, \label {eq40a}\\
\tilde{X} (\zeta,k)&=&c (1-k^{2}-2ik\tanh\zeta-2sech^{2}\zeta) e^{ik\zeta}. \label {eq40b}
\end{eqnarray}
\end{subequations}
On comparing Eqs.~(\ref{eq40a}) and (\ref{eq40b}) we can write
\begin{eqnarray}
\tilde{X} (\zeta,k)=\frac{{X_{\zeta}(\zeta,k)}}{ik}. \label{eq41}
\end{eqnarray} 
Using Eq.(\ref{eq41}) in the right hand side of Eq.(\ref{eq29}) and comparing
 the resultant equation with  (\ref{eq28a}), we can write down the eigen value 
$\lambda_{0}$ as
\begin{eqnarray}
\lambda_{0}=i\frac{(1+k^{2})}{k}.\label{eq42}
 \end{eqnarray} 
\\

 Now to determine the constant `$c$' we use the orthonormality relation between
 $X(\zeta,k)$ and $\tilde {X}(\zeta,k)$ given by
\begin{eqnarray}
\int_{-\infty}^{\infty} X(\zeta,k) {\tilde{X}} ^{\ast} (\zeta,k') d\zeta=\delta
 (k-k'). \label{eq43}
\end{eqnarray}
 On substituting the eigen functions $X(\zeta,k)$ and $\tilde {X}(\zeta,k)$ given
  in Eqs.~(\ref{eq40a}) and ~(\ref{eq40b}) respectively in Eq.(\ref{eq43})
  and after evaluating the integral we obtain 
 \begin{eqnarray}
 2\pi c^{2} \left[ (1-k^{2}) (1-k'^{2})+4 kk'\right] \delta (k-k')=\delta
 (k-k'), \label{eq44}
 \end{eqnarray}
 which gives $c^{2}=\frac{1}{2\pi}(1+k^{2})^{-2}$. The correct form of the eigen
 functions $X(\zeta,k)$ and $\tilde {X}(\zeta,k)$ is written down after using
 the above value of `$c$' in 
 Eqs.~(\ref{eq40a}) and (\ref{eq40b}).
\begin{subequations} 
\begin{eqnarray}
X(\zeta,k)&=&\frac{(1-k^{2}-2ik\tanh\zeta)}{\sqrt{2\pi}(1+k^{2})}e^{ik\zeta}  ,
\label{eq45a}\\
\tilde{X} (\zeta,k)&=&
\frac{(1-k^{2}-2ik\tanh\zeta-2sech^{2}\zeta)}{\sqrt{2\pi}(1+k^{2})} e^{ik\zeta} .
 \label{eq45b}
\end{eqnarray}
\end{subequations}
 It  can  be  verified  that the operator $L_{2}$ also has the following two
 discrete eigen functions.
\begin{eqnarray}
\tilde{X}_{0} (\zeta)=(1-\zeta\tanh\zeta)sech\zeta ,\quad
\tilde{X}_{1} (\zeta)=sech\zeta \tanh\zeta. \quad\label{eq46}
\end{eqnarray}
It may be further noted that the eigen function $\tilde {X}_{1}$ corresponds
 to the discrete eigen value $\lambda=0$. That is
\begin{eqnarray}
 L_{2} \tilde{X}_{1}(\zeta)=0. \label{eq47}
\end{eqnarray}
\subsection{Complete set of orthonormal basis}
 Having found the eigen functions $X(\zeta,k)$ and $\tilde {X}(\zeta,k)$, we now
 check the completeness of them
  by writing 
\begin{eqnarray}
\int_{-\infty} ^{\infty} X(\zeta,k) {\tilde{X}}^{\ast} (\zeta',k) dk+f(\zeta,\zeta')=
\delta (\zeta-\zeta'), \label{eq48}
\end{eqnarray}
where $f(\zeta,\zeta')$ is an arbitrary function to be determined.
First we evaluate the integral in the left hand side  of Eq.~(\ref{eq48}) after 
substituting the values of the eigen
functions $X(\zeta,k)$ and $ \tilde{X} (\zeta,k)$ given in Eqs.~(\ref{eq45a}) 
and (\ref{eq45b}). Thus we have the following integrals to evaluate.
\begin{eqnarray}
\int_{-\infty} ^{\infty} X(\zeta,k) {\tilde{X}}^{\ast} (\zeta',k) dk&=&\delta
(\zeta-\zeta') 
-\frac{1}{\pi}\left[ \int_{-\infty}^{\infty} 
\frac{dk}{(1+k^{2})}e^{ik(\zeta-\zeta')}  \right.\nonumber\\
&& \times\{2-sech^{2}\zeta'-ik(\tanh\zeta -\tanh\zeta')
\}\nonumber\\
&&-2\int_{-\infty}^{\infty}\frac{dk}{(1+k^{2})^{2}}
 e^{ik(\zeta-\zeta')}\tanh\zeta'\nonumber\\
 &&\times(1-ik\tanh\zeta)
\left.(ik+\tanh\zeta')\right].\label{eq49}
\end{eqnarray}
The  integrals in the right hand side of Eq.(\ref{eq49}) are evaluated
with the aid of the residue theorem. It
may be noted that the integrands in these two integrals as functions of the
complex variable $k$ are analytic everywhere in the complex $k$-plane except at the two
poles $k=\pm i$ of first and second order respectively. Let 
$R_{1}$ and $R_{2}$ be  the residues corresponding to the functions
 $\frac{1}{(1+k^{2})}e^{ik(\zeta-\zeta')}
 \{2-ik(\tanh\zeta -\tanh\zeta')
-sech^{2}\zeta'\}$ and $\frac{1}{(1+k^{2})^{2}}
 e^{ik(\zeta-\zeta')}(1-ik\tanh\zeta)
(ik+\tanh\zeta')\tanh\zeta'$ in Eq.~(\ref{eq49}) at  the pole $k=i$ of first and second order 
 respectively. On calculating the residues $R_1$ and $R_2$ using  standard procedure we obtain
\begin{subequations}
\begin{eqnarray}
R_1&=&\frac{-i}{2}
\left[2+\tanh\zeta-\tanh\zeta'-sech^{2}\zeta\right]
e^{(\zeta-\zeta')},\label{eq50a}\\
R_2&=&\frac{-i}{4}
\left[(1-\zeta+\zeta')(\tanh\zeta-\tanh\zeta')-(\zeta-\zeta')\right.\nonumber\\
 &&\left.\times(1-\tanh\zeta\tanh\zeta')\right] e^{(\zeta-\zeta')}.
\label{eq50b}
\end{eqnarray}
\end{subequations}
On summing up the above results Eq.~(\ref{eq49}) becomes
\begin{eqnarray}
\int_{-\infty}^{\infty} X(\zeta,k){\tilde{X}}^{\ast} (\zeta',k) dk &=& \delta
 (\zeta-\zeta')-\left[sech\zeta sech\zeta'(1-\zeta'\tanh\zeta')\right.\nonumber\\
 &&\left.+\zeta sech\zeta sech\zeta'\tanh\zeta'\right],\label{eq51}
\end{eqnarray}
which can be identified as
\begin{eqnarray}
\int_{-\infty}^{\infty} X(\zeta,k){\tilde{X}}^{\ast} (\zeta',k) dk=\delta
 (\zeta-\zeta')
-\left[X_{0} (\zeta){\tilde{X}}_{0} (\zeta')+X_{1} (\zeta) {\tilde{X}}_{1}
(\zeta') \right],\label{eq52}
\end{eqnarray}
upon using the values of  $\tilde {X}_{0}(\zeta)$ and $\tilde{X}_{1}(\zeta)$
 as given in Eq.(\ref{eq46}). By 
substituting Eq.(\ref{eq52}) in Eq.(\ref{eq48}), we can evaluate the  
following value  of $f(\zeta,\zeta')$ in terms of the discrete 
orthogonal  states.
\begin{eqnarray}
f(\zeta,\zeta')=\sum_{j=0,1} X_{j}(\zeta) {\tilde{X}}_{j} (\zeta). \label{eq53}
\end{eqnarray}
Thus,  by comparing Eqs.(\ref{eq51}) and (\ref{eq52}), we can write two additional
  orthogonal  discrete  states given by
\begin{eqnarray}
X_{0} (\zeta)= sech\zeta ,\quad
X_{1} (\zeta)= \zeta sech\zeta . \label{eq54}
\end{eqnarray}
It may be checked that the following relations exist between 
the  discrete states $X_{0}(\zeta),\tilde {X}_{0}(\zeta),X_{1}(\zeta)$ and
$\tilde {X}_{1}(\zeta)$.
\begin{eqnarray}
L_{1} X_{0}=0,~L_{1}X_{1}(\zeta)= -2\tilde{X}_{1}(\zeta),~ L_{2} \tilde{X}_{0}(\zeta)=2
X_{0}(\zeta), \label{eq55}
\end{eqnarray}
in addition to the relation given in Eq.(\ref{eq47}).
In conclusion, we have a set of two complete orthonormal
 bases $\{X\}$ and $\{\tilde{X}\}$ given
by $\{X(\zeta,k),X_{0}(\zeta),X_{1}(\zeta)\}$ and
$\{\tilde{X}(\zeta,k),\tilde{X}_{0}(\zeta),\tilde{X}_{1}(\zeta)\}$ respectively.
 These set of orthonormal basis functions
will be used to construct the perturbed soliton solution.
\subsection{Evaluation of $T(\tau)$}
Having solved Eq.(\ref{eq28a}) by finding  $X(\zeta)$, in order to construct
 $\psi^{(1)}(\zeta,\tau)$,
 we now find $T(\tau)$ by solving
Eq.~(\ref{eq28b}). For this first we rewrite Eq.~(\ref{eq28b}) by replacing 
the function $H(\tau)$ in the right hand side
 using Eq.~(\ref{eq26}) and
(\ref{eq41}).
\begin{eqnarray}
T_{\tau}-\lambda_{0} T=\frac{F^{(1)}(\zeta,\tau)}{i k \tilde{X}(\zeta,k)}. \label{eq56}
\end{eqnarray}
 On multiplying and dividing the right hand side of Eq.~(\ref{eq56})
 by ${X}^{\ast}(\zeta,k)$ and on integrating  with respect to $\zeta$
 between the limits $-\infty$ to $+\infty$ we obtain due to orthonormality
 of the functions $X$ and $\tilde{X}$  (see Eq.(\ref{eq43})) the following equation.
\begin{eqnarray} 
T_{\tau}-\lambda_{0} T=\frac{1}{ik} \int_{-\infty}^{\infty} F^{(1)}(\zeta,\tau)
{X}^{\ast}(\zeta,k) d\zeta, \label{eq57}
\end{eqnarray}
which can be explicitly written  after substituting the values of
$F^{(1)}(\zeta,\tau)$ and $X^{\ast}(\zeta,k)$ from  Eqs.~(\ref{eq24b}) and (\ref{eq45a})
respectively as 
\begin{eqnarray}
T_{\tau}-\lambda_{0} T&=&\frac{2}{i\sqrt{2\pi} k (1+k^2)} \int_{-\infty}^{\infty}d\zeta
\left(\left[g(\zeta) sech\zeta\right]_{\zeta} \right.
+2v_0~sech\zeta\left[m_{t_{1}}\right.\nonumber\\
&&\left.\left.+(m^{2}\xi_{t_{1}}-\zeta
m_{t_{1}})\tanh\zeta
 \right]\right) (1-k^{2}+2ik\tanh\zeta)e^{-ik\zeta}.\label{eq58}
\end{eqnarray}
On solving the above equation using standard procedure, we obtain
\begin{eqnarray}
T(\tau,k)=C(k) e^{\lambda_{0}\tau}-\frac{2}{i \sqrt{2\pi}\lambda_{0}k(1+k^{2})}
\int_{-\infty}^{\infty}  d\zeta 
\left(\left[g(\zeta) sech\zeta\right]_{\zeta}+2v_0~sech\zeta\right.\nonumber\\
\left.\times\left[m_{t_{1}}+(m^{2}\xi_{t_{1}}-\zeta
m_{t_{1}}) \tanh\zeta
 \right]\right)(1-k^{2}+2ik\tanh\zeta)e^{-ik\zeta},\qquad
\label{eq59}
\end{eqnarray}
where C(k) is a constant which can be found by using the initial
condition $T(\tau,k)=0$ when $\tau\rightarrow
-\frac{(1+v)\zeta}{2}$. Thus we obtain
\begin{eqnarray}
C(k)&=&\frac{2}{i \sqrt{2\pi}\lambda_{0}k(1+k^{2})}\int_{-\infty}^{\infty} d\zeta
\left(\left[g(\zeta) sech\zeta\right]_{\zeta}\right.
+2v_0~sech\zeta\left[m_{t_{1}}\right.
\nonumber\\
&&\left.\left.+(m^{2}\xi_{t_{1}}-\zeta
m_{t_{1}})\tanh\zeta
 \right]\right) (1-k^{2}+2ik\tanh\zeta)e^{\lambda_{0}(1+v_{0})\frac{\zeta}{2}-ik\zeta},\qquad
\label{eq60}
\end{eqnarray}
and hence
\begin{eqnarray}
T(\tau,k)&=&\frac{2}{\sqrt{2\pi}i\lambda_{0} k(1+k^{2})}\int_{-\infty}^{\infty}
d\zeta' \left(\left[g(\zeta') sech\zeta' \right]_{\zeta'}\right.
+2v_0~sech\zeta'\left[m_{t_{1}}\right.\nonumber\\
 &&\left.\left.+(m^{2}\xi_{t_{1}}-\zeta'
 m_{t_{1}})\tanh\zeta' \right] \right)
  (1-k^{2}+2ik\tanh\zeta')  e^{-ik\zeta'}\nonumber\\
&& \times\left(e^{\lambda_{0}
[\tau+\frac{(1+v_{0})}{2}\zeta']}-1\right).
 \label{eq61}
\end{eqnarray}
The allowed discrete values of $T$ will be determined in the next section while
constructing the perturbed part of the soliton 
$\psi^{(1)}(\zeta,\tau)$ using the values of  $X(\zeta)$ and $T(\tau)$.
\subsection{Perturbation of soliton}
 As mentioned, we now write down the first order perturbation correction
 $\Psi^{(1)}(\zeta,\tau)$ in terms of the basis functions 
$\{X\}\equiv    \{X(\zeta,k),X_{0}(\zeta),X_{1}(\zeta)\}$ and
 $ \{T\}\equiv\{T(\tau,k),T_{0}(\tau),T_{1}(\tau)\}$. As per Eq.(\ref{eq26}),
  in terms of the basis functions the
 perturbed part of the soliton can be written as
\begin{eqnarray}
\Psi^{(1)}(\zeta,\tau)=\int_{-\infty}^{\infty} X(\zeta,k) T(\tau,k)
dk+\sum_{j=0,1}X_{j} (\zeta) T_{j}(\tau). \label{eq62}
\end{eqnarray}
 However, it should be noted that in the basis $\{T\}, T_{0}(\tau)$  and
 $T_{1}(\tau)$ are yet to be determined. Therefore before 
 evaluating the values of $\Psi^{(1)}(\zeta,k)$,  
 we determine $T_{0}(\tau)$ and $T_{1}(\tau)$. This is carried out 
 by substituting Eq.(\ref{eq62}) in
 Eq.(\ref{eq25})  which finally becomes
\begin{eqnarray}
\int_{-\infty}^{\infty} ik\left[ T_{\tau} (\tau,k)-\lambda_{0} T(\tau,k)\right]\tilde{X}(\zeta,k) 
dk+{T_{1}}_{\tau} (\tau)\tilde{X}_{0} (\zeta)\nonumber\\
 -\left[ {T_{0}}_{\tau} (\tau)-2 T_{1} (\tau)\right]
 \tilde{X}_{1}(\zeta)=F^{(1)}(\zeta,\tau),  \label{eq63}
\end{eqnarray}
  upon using Eqs.(\ref{eq29}),(\ref{eq41}),(\ref{eq47}) and (\ref{eq55}). Now multiplying
  Eq.(\ref{eq63}) by $X^{\ast}(\zeta,k), X_{0}(\zeta)$ and $X_{1}(\zeta)$ 
  separately and using the orthonormal relations such as     
 $ \int_{-\infty}^{\infty}\tilde{X}(\zeta,k)
 X^{\ast}(\zeta,k) dk=1,\int_{-\infty}^{\infty} \tilde{X}(\zeta,k) X_{j}(\zeta)
 d\zeta\equiv\int_{-\infty}^{\infty}X(\zeta,k)\tilde{X}_{j}(\zeta)d\zeta=0,
  \int_{-\infty}^{\infty}X_{j}(\zeta)\tilde{X}_{l}(\zeta)
 d\zeta=\delta_{jl}, j,l=0,1$, we get
  
\begin{subequations}
\begin{eqnarray}  
  T_{\tau}(\tau,k)-\lambda_{0} T(\tau,k)&=&\frac{1}{ik} \int_{-\infty}^{\infty} F^{(1)}
 (\zeta,\tau)X^{\ast} (\zeta,k) d\zeta, \label{eq64a}\\
{ T_{1}}_{\tau} (\tau)&=&\int_{-\infty}^{\infty} F^{(1)} (\zeta,\tau) X_{0} (\zeta)
d\zeta , \label{eq64b}\\
 {T_{0}}_{\tau} (\tau)-2T_{1} (\tau)&=&-\int_{-\infty}^{\infty} F^{(1)}
 (\zeta,\tau)
X_{1} (\zeta) d\zeta.  \qquad\label{eq64c}
\end{eqnarray}
\end{subequations}
As $F^{(1)}(\zeta,\tau)$  given in Eq.~(\ref{eq24b}) does not contain time
`$\tau$'
explicitly, the right
hand side of Eqs.~(\ref{eq64a}-c) also should be independent of time. Then it is
easy to verify that for the values of  $F^{(1)}(\zeta,\tau)$,   
$X^{\ast}(\zeta,k), T(\tau,k)$   and $ \lambda_{0}$ given respectively in Eqs.
(\ref{eq24b}), (\ref{eq45a}), (\ref{eq61}) and (\ref{eq42}), Eq.~(\ref{eq64a})
satisfies identically. As the right hand sides of Eqs.~(\ref{eq64b}) and
~(\ref{eq64c}) are also independent of time, they give rise to secularities and
hence the nonsecular conditions can be written as
\begin{subequations}
\begin{eqnarray}
\int_{-\infty}^{\infty}F^{(1)} (\zeta,\tau) X_{0} (\zeta) d\zeta=0, \label{eq65a}\\
\int_{-\infty}^{\infty}F^{(1)} (\zeta,\tau) X_{1} (\zeta) d\zeta=0. \label{eq65b} 
\end{eqnarray}
\end{subequations}
Using Eqs.~(\ref{eq65a}) and (\ref{eq65b}) back in Eqs.~(\ref{eq64b}) and ~(\ref{eq64c}),
 we obtain 
 $T_{1}(\tau)=0$ and $\quad T_{0} (\tau)=C_{1}$  which has  to be determined. For
 this, we substitute  $T_{1}(\tau)=0$ in Eq.~(\ref{eq64c}) and
 integrate with respect to $\tau$ to obtain 
 \begin{eqnarray}
T_{0}(\tau)\equiv C_{1}&=&(1+v)\int_{-\infty}^{\infty}d\zeta\left(\left[g(\zeta) sech\zeta \right]_{\zeta}+2v_0~sech\zeta\right.\nonumber\\
&&\times\left.\left[m_{t_{1}}+(m^{2}\xi_{t_{1}}-\zeta
m_{t_{1}})\tanh\zeta \right]\right)\zeta^{2} sech\zeta.\quad
  \label{eq66}
\end{eqnarray}
\subsection{Variation of soliton parameters}
We now estimate the  nonsecularity conditions (\ref{eq65a}) and (\ref{eq65b})
 by evaluating the
integrals  after substituting the
values of $F^{(1)}(\zeta,\tau), X_{0}(\zeta)$ and $X_{1}(\zeta)$ respectively
 from Eqs.(\ref{eq24b}) and (\ref{eq54}). The results give
\begin{subequations}
\begin{eqnarray}
m_{t_{1}}&=&-\frac{1}{2v_0} \int_{-\infty}^{\infty} \left[ {g(\zeta) sech\zeta}\right]
_{\zeta} sech\zeta d\zeta,  \label{eq67a}\\
\xi_{t_{1}}&=&-\frac{1}{2m^{2} v_0}\int_{-\infty}^{\infty} \left[{g(\zeta) sech\zeta}\right]
_{\zeta}   \zeta sech\zeta d\zeta . \label{eq67b}
\end{eqnarray}
\end{subequations}
Eq.~(\ref{eq67a}) describes the time evolution of the inverse of the
 width  of the soliton and
Eq.~(\ref{eq67b}) gives the velocity of the soliton. $g(\zeta)$ that appears in
the above nonsecularity relation 
is related to the inhomogeneity in stacking energy of DNA. In order to evaluate
the integrals in Eqs.~(\ref{eq67a}) and (\ref{eq67b}) explicitly
  we have to substitute  specific value for $g(\zeta)$. 
We consider  $g(\zeta)$ in the form of  
localized and 
periodic functions. A localized  $g(\zeta)$ corresponds to the intercalation of a
compound between neighbouring base pairs without disturbing the base pairs 
 and their sequence in the
DNA double helical chain. The periodic nature of $g(\zeta)$  represents a
 periodic
repetition of similar base pairs along the helical chain. 
We consider the localized form of $g(\zeta)$ as (i)
$g(\zeta)=sech\zeta$  and the periodic form 
of $g(\zeta)$ as (ii) $g(\zeta)=\cos\zeta$.  We substitute the above values of $g(\zeta)$ 
one by one in Eqs.~(\ref{eq67a}) and 
(\ref{eq67b}) and evaluate the integrals in the right hand side to understand the time evolution of the width
of the soliton and its velocity. At this point it is worth mentioning that Dandoloff and Saxena \cite{ref36}
realized that in the case of XY-spin chains the model of which identifies with our plane-base rotator model,
the ansatz $g(\zeta)=sech\zeta$ energetically favours the deformation of spin chains.\\

 When we substitute $g(\zeta)=sech\zeta$ in Eqs.~(\ref{eq67a}) and (\ref{eq67b})
and on  evaluating the integrals, we obtain
\begin{eqnarray}
m_{t_{1}}=0, \quad \xi_{t_{1}}=\frac{\pi}{6m^{2}v_{0}}. \label{eq68}
\end{eqnarray}
The above equations can be rewritten in terms of the original time variable $ \hat t $ 
by using the transformation
  $\frac{\partial }{\partial \hat t}=\frac{\partial }{\partial t_0}+\epsilon
\frac{\partial }{\partial t_1}$ or in otherwords 
$m_{\hat t}=m_{t_{0}}+\epsilon m_{t_{1}}$ and $ \quad \xi_{\hat t}=\xi_{t_{0}}+\epsilon
\xi_{t_{1}}.$ As m is independent of $t_{0} (m_{t_{0}}=0)$ and $\xi_{t_{0}}=v_{0}$, we can write
\begin{eqnarray}
m=m_{0}, \quad \xi_{\hat{t}}\equiv v=v_{0}+\frac{\epsilon\pi}{6m_{0}^{2}v_{0}}, \label{eq69}
\end{eqnarray}
where $1/m_{0}$ is the initial width of the soliton. The first of  Eq.(\ref{eq69}) 
says that when $g(\zeta)=sech\zeta$, the  width $(m^{-1})$ 
of the soliton remains constant. However from the second of Eq.(\ref{eq69}), we
find that the velocity of the 
soliton gets a correction. As the correction term is a definite positive quantity the velocity of the soliton
increases in this case. Interestingly, the amount of increment in velocity depends on the initial width and
initial velocity
of the soliton. Wider the soliton, greater the increment in velocity due to inhomogeneity. Also slowly moving solitons gain more
speed. The increase in speed helps to overcome the barrier of the local inhomogeneity (which may
be due to the presence of abasic site or intercalation of  a molecule) and the
solitons representing the open state will propagate easily along the chain without formation of a
bound state. In a  similar study on resonant kink-impurity interaction  and kink scattering in the sine-Gordon model 
 Zhang Fei et al. \cite{ref32}  observed
that if the  initial velocity of the kink is smaller than a critical velocity  it will be either trapped or reflected
by the  impurity.  In fact they have showed that for most of the initial velocities, the kink is trapped except in the case of some special initial velocities  the kink may be totally reflected by the  impurity. It was further found that when the kink velocity is greater than the critical velocity, it will pass through the impurity. It is interesting to note that our results  on the velocity of soliton found in Eq. (\ref{eq69})  is similar to the last case of Zhang Fei et al \cite{ref32} where the soliton will pass through  by overcoming the barrier of the local inhomogeneity.\\
 
 Next, we substitute 
  the periodic function $g(\zeta)=\cos\zeta$  in
 Eqs.(\ref{eq67a}) and (\ref{eq67b})
and evaluate the integrals to obtain
\begin{eqnarray}
m_{t_{1}}=0, \quad
\xi_{t_{1}}=\frac{\pi^{2}}{16 m^{2}v_{0}}.\label{eq72}
\end{eqnarray}
 From the above the parameters m and $\xi$ can be written in terms of the original
 variable $\hat t$  after solving Eq.(\ref{eq72}) as
\begin{eqnarray}
m=m_{0}, \quad
\xi_{\hat {t}}\equiv v=v_{0}+\frac{\epsilon\pi^{2}}{16 m^{2} v_{0}}.\label{eq73}
\end{eqnarray}
 From 
Eq.~(\ref{eq73}) we
find that the  width of the soliton in this case remains 
 constant and the soliton velocity  increases  similar to the case
   $g(\zeta)=sech\zeta$. On comparing Eqs.~(\ref{eq69}) and~(\ref{eq73}),
  we observe that the increase in velocity
  of the soliton is less in this case. This is because in this case,  the
  inhomogeneity occurs periodically in the entire DNA chain in terms of sequence.   
\subsection{First order perturbed soliton}
Now, we explicitly construct the first order perturbation correction to the one  soliton
 for the different cases of $g(\zeta)$
 by substituting the values of the basis functions $\{X\}
\equiv\{X(\zeta,k), X_{0}(\zeta), X_{1}(\zeta)\}$ and
 $\{T\}\equiv\{T(\tau,k), T_{0}(\tau), T_{1}(\tau)\}$  after  using the values
  of $F^{(1)}(\zeta,\tau), m_{t_{1}}$ and $\xi_{t_{1}}$ for the
  respective $g(\zeta)$ values  in Eq.~(\ref{eq62}).
 Thus in the case of  $g(\zeta)=sech\zeta$, we substitute the values of
   $X(\zeta,k),
X_{0}(\zeta)$ and $ X_{1}(\zeta)$ from  Eqs.~(\ref{eq45a}) and
(\ref{eq54}) and $T(\tau,k),~T_{0}(\tau)$ from Eqs.~(\ref{eq61}) and
(\ref{eq66}) and $F^{(1)}(\zeta,\tau)$  from Eq.~(\ref{eq24b}) and use the
values of $m_{t_{1}}$ and $\xi_{t_{1}}$ from Eqs.~(\ref{eq68}) in Eq.~(\ref{eq62}) to obtain
\begin{eqnarray}
\Psi^{(1)} (\zeta,t_{0})&=&-\frac{1}{3\pi}\int_{-\infty}^{\infty}
\frac{dk}{(1+k^{2})^{3}} (1-k^{2}-2ik\tanh\zeta)
e^{ik\zeta}\int_{-\infty}^{\infty}d\zeta\nonumber\\
&&\times(1-k^{2}+2ik\tanh\zeta)
 (\pi-6sech\zeta) sech\zeta\tanh\zeta \nonumber\\
&&\times\left[e^{-ik\zeta} -e^{i\frac{(1+k^{2})}{k}\alpha}e^{i\beta\zeta}\right],
  \label{eq74}
\end{eqnarray}
where $\alpha=\frac{t_{0}-m(1+v_{0})\zeta}{2m}$ 
and $\beta={\frac{(1+v_{0})(1+k^{2})}{2k}}-k$. 
While writing the above, we have also used $T_{1}(\tau)=0, \tau=\frac{1}{2m}[t_{0}-m(1+v_{0})\zeta]$ and the eigen value
$\lambda_{0}=\frac{i(1+k^{2})}{k}$. 
The integrals in the right hand side of Eq.(\ref{eq74}) can be evaluated using 
 standard residue
theorem . The details are given in Appendix-A.
The final form of $\Psi^{(1)}(\zeta,t_{0})$ after evaluating the integrals  
 becomes
\begin{eqnarray}
\Psi^{(1)}(\zeta,t_{0})&\approx&\frac{80}{27\sqrt{2}}\sqrt{sech\zeta}~e^{-\frac{3\zeta}{2}}
 +\frac{64}{27\sqrt[3]{2}}\tanh\zeta
 \sqrt[3]{sech\zeta}~e^{-\frac{5}{3}\zeta}\nonumber\\
&&  +\frac{\pi}{6v^{2}} 
\left[2 v(1+v)\zeta
+v^{2}+4\alpha v-1\right]sech\zeta.  \label{eq75}
\end{eqnarray}
While constructing the above perturbed solution, we have used the
 values of m's and $\xi$'s  and of course the
corresponding $F^{(1)}(\zeta,t_0)$ values.
Finally,  the perturbed one soliton solution  that is
$\Psi(z,t_{0})=\Psi^{(0)}(z,t_{0})+
\Psi^{(1)}(z,t_{0})$ (choosing $\epsilon=1$) is written 
using Eqs.~(\ref{eq22}) and (\ref{eq75}) as
 \begin{figure}
\begin{center}
\epsfig{file=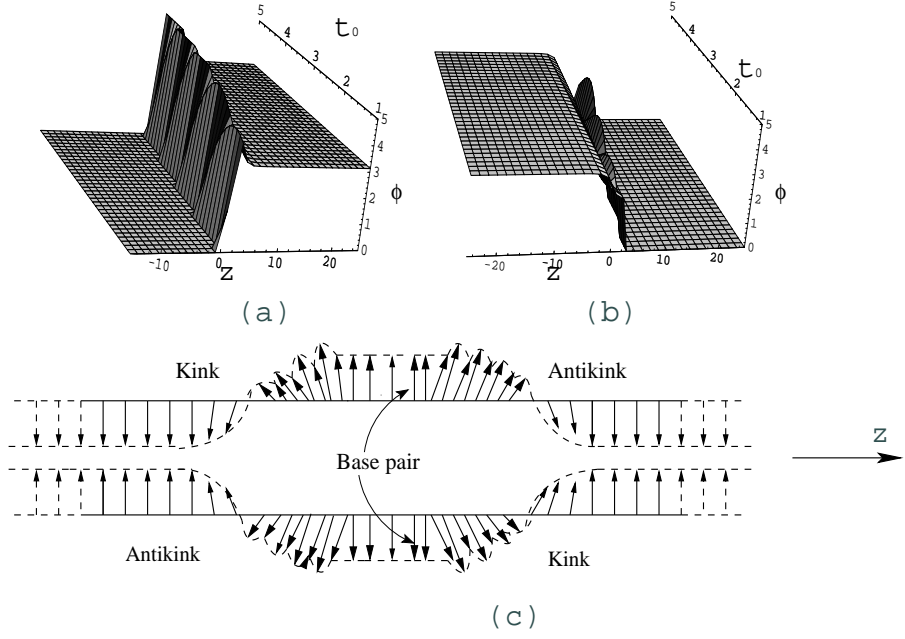,height =10cm, width=12cm}
\caption{(a) The perturbed kink-soliton and (b) the perturbed antikink-soliton for the inhomogeneity
$g( z)=sech z$ and $v_0=0.4$. (c) A  sketch of the  base pair opening in DNA double helix  with fluctuation.}
\end{center}
\end{figure} 
\begin{eqnarray}
\Psi(z,t_{0})&\approx& 4
arc\tan\exp[\pm m_{0}(z-v_{0}t_{0})]+\frac{80}{27\sqrt{2}}
\sqrt{sech[\pm m(z- v t_{0})]}\nonumber\\
&\times& e^{\mp\frac{3(m(z-v t_{0})}{2}}
+\frac{64}{27\sqrt[3]{2}}\tanh[\pm m(z-v t_{0})]
 \sqrt[3]{sech[\pm m(z-v t_{0})]} \nonumber\\ 
&\times& e^{\mp\frac{5}{3}m(z-v t_{0})}+\frac{\pi}{6 m v^{2}} 
\left[m (v^{2}-1)+2 t_{0} v\right]sech[\pm m(z-v t_{0})].  \label{eq76}
\end{eqnarray}
 In Eq.(\ref{eq76}) while the upper sign corresponds to perturbed kink-soliton   the lower sign represents
 the perturbed antikink-soliton.
  The rotation of bases  denoted by $\phi(z,t_{0})$  can be immediately written down by using the
   relation $\phi=\frac{\Psi}{2}$.
In Figs. 4a and 4b, we plot  $\phi(z,t_{0})$ (rotation of bases under perturbation)  for the
 parametric choice
 $v_{0}=0.4$. From the figure, we observe that  there appears fluctuation in the form of
 a train of pulses closely resembling the shape of the inhomogeneity profile in the width of the soliton as
 time progresses. Also, as time passes, the amplitude  of the  pulses generating this fluctuation
 increases. However, in the asymptotic region of the soliton there is no change in the topological character
 and no fluctuations appear in that region. It shows that 
the localized inhomogeneity  in stacking energy  in DNA in the form of a pulse ($g(z)=sech z$) does not 
affect very much the opening of bases except 
fluctuations in the form of a train of pulses in the localized region of the kink and antikink-soliton.
 We have schematically represented this in  Fig.4c.\\

We  then repeat the procedure  for constructing the perturbed one-soliton solution in the case of $g(\zeta)=\cos\zeta$ which  is written as
\begin{eqnarray}
\Psi^{(1)} (\zeta,t_{0})&\approx&\frac{1}{\pi}\int_{-\infty}^{\infty}
\frac{dk}{(1+k^{2})^{3}} (1-k^{2}-2ik\tanh\zeta) e^{ik\zeta} \int_{-\infty}^{\infty}
d\zeta(1-k^{2}\nonumber\\
&& +2ik\tanh\zeta)[\sin\zeta+(\cos\zeta -\frac{\pi^{2}}{8})\tanh\zeta]sech\zeta
 \nonumber\\
&&\times\{e^{i\frac{(1+k^{2})}{k}\alpha}e^{i\beta\zeta}-e^{-ik\zeta}\}
 +(1+v)sech\zeta\nonumber\\
&& \times\int_{-\infty}^{\infty}
d\zeta ~\zeta^{2} [\sin\zeta+(\cos\zeta-\frac{\pi^{2}}{8})\tanh\zeta] sech^{2}\zeta .\label{eq79}
\end{eqnarray}
The details of  values of the integrals in the above equation are  given in
Appendix-B.    
The perturbed  kink (upper sign)-antikink (lower sign) one soliton solution in this case   is finally 
obtained as
 \begin{eqnarray}
 \Psi(z,t_{0})&\approx&4~
arc\tan\exp[\pm m_{0}(z-v_{0} t_{0})]
+\frac{\pi^{2}}{16 m v^{2}}\left[m(v^{2}-1)+2 v t_{0}\right]\nonumber\\
&& \times sech[ m(z-v t_{0})].\label{eq80}
 \end{eqnarray}
After finding  $\phi(z,t_{0})$ from the relation $\phi=\frac{\Psi}{2}
$ we plot  it in Figs.5a and 5b for the same value of the parameter as before. From the figures we observe
that periodic oscillations appear  in the width of the soliton without any change
asymptotically.
  \begin{figure}
\begin{center}
\epsfig{file=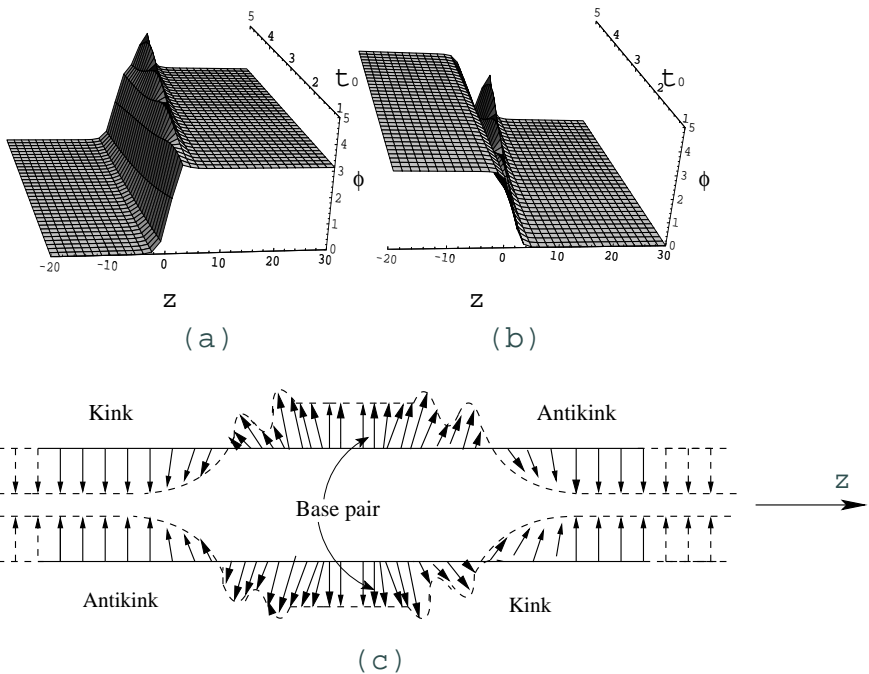,height =10cm, width=12cm}
\caption{ The perturbed (a) kink-soliton and (b)  antikink-soliton for the inhomogeneity
$g(z)=\cos z $ with $v_{0}=0.4$. (c) A  sketch of the open state configuration in DNA  
 with small fluctuations.}
\end{center}
\end{figure}  
\section{Conclusions }
 In this paper, we 
  studied the nonlinear dynamics of DNA  double helix with stacking inhomogeneity
   by considering the
 dynamic plane-base rotator
 model. The  dynamical equation which finally appeared in the form of a perturbed
 sine-Gordon equation was derived from a suitable Hamiltonian in analogy with Heisenberg model of
  an inhomogeneous anisotropic coupled spin chain or spin ladder with ferromagnetic legs and antiferromagnetic
 rung coupling in the continuum limit. In the unperturbed  limit
  which is also the homogeneous limit, the dynamics is governed by the kink-antikink soliton of the integrable sine-Gordon
 equation which represents the open state configuration of base pairs in DNA double helix. 
  Even though DNA double helix is a large molecular chain involving very large number of base
  pairs,  base pair opening is limited to a very few number of base pairs forming
  localized coherent structure in the form of kink-antikink solitons which is obtained as a balance
  between dispersion and nonlinearity traveling with constant speed and amplitude without 
  loosing its energy along the
  helical chain. To understand the
 effect of stacking energy inhomogeneity  on the open state configuration of base pairs we carried out a
 multiple scale soliton perturbation analysis. For implementing this  we linearized 
 the perturbed sine-Gordon equation  using multiple-scale expansion
  to obtain linearized equations  in the form of eigen value problems. The perturbed kink-antikink
  soliton solutions were constructed for  different forms of inhomogeneities by solving the   associated eigen value
  problem. By using the complete set of eigen functions thus obtained as the basis functions  the perturbed
  solutions were constructed.\\
  
   The perturbation
 not only modifies the shape of the soliton but also undergoes a slow time
 change of the soliton parameters namely the width and velocity for the different forms
  of the inhomogeneity chosen. We chose the inhomogeneity in the form of localized and
  periodic functions. The results show that when the inhomogeneity is either in the form 
  $g(z)=sech z$ or $g(z)=\cos z$, the width of the soliton remains constant.  Thus, the number
   of base pairs participating in the opening do not change due to the above pattern of
   inhomogeneities. However, in this case  the
  speed of the soliton increases with a correction that is proportional to the square of the
  initial width of the soliton and inversely proportional to its initial velocity. Thus the inhomogeneity increases the speed with which the base pairs
  are opening and closing or winding and unwinding.
   From the perturbed solutions  corresponding to different inhomogeneities  (see Figs. 4,5) we  observe
   that the perturbation due to
   inhomogeneity  in stacking energy along the strands introduces  fluctuation only in the width of the
   solitons. The nature of the  fluctuation varies depending on the type of  inhomogeneity. 
   In particular, when the inhomogeneity is in the form of  $g(z)=sech z$, we find that fluctuation in the form of
   pulse trains resembling the shape of the inhomogeneity is generated in the width of the soliton
   representing open state configuration.  It is noted that in the long time limit, eventhough
   neighbouring pulses overlap the train of pulse-like fluctuations maintain their character without
   affecting the soliton. In a similar way, when the inhomogeneity is of the form
   $g(z)=\cos z$, the fluctuation appears in the form of periodic oscillations in the width of the
    soliton.  In all these cases, asymptotically the kink-antikink soliton shape is preserved.
   The results indicate that inhomogeneity in stacking energy in  DNA double helix can  (i) introduce small
   fluctuations which may merge asymptotically during the process of opening and closing of base pairs (ii) increase or decrease the number of base pairs
   participating in the open state configuration and (iii) change the speed with which the open state
   configuration can travel along the double helical chain. Thus in conclusion the inhomogeneity in
   stacking does not affect the general pattern of base pair opening. Even though the size of  the base pair
   is big and the solvent effect on pure rotation of base pairs is negligible, the soliton,
   a coherent structure which is formed by  involving few base pairs move along the helical chain
   without dissipation or any other form of deformation.  Similar conclusion was also arrived in
   the case of propagation of soliton representing base pair opening in a discrete site-dependent
   DNA \cite{ref24} and propagation of bubble in a heterogeneous DNA chain \cite{ref20}. As nature selects generally inhomogeneous DNA,
   the functions such as replication and transcription can be explained more viably through formation of open
   states  through our inhomogeneous model rather than the homogeneous  case. This is also because,
   it is known that transcription and replication are sequence dependent.  This is further similar to
   what was observed in the case of proteins where inhomogeneity of the sequence leads to
   inhomogeneous fluctuations, enhanced by the nonlinear effect \cite{ref50}. Eventhough the
   relevance of these effects in the DNA to biological processes are not yet clearly, established
   a recent study suggests that thermally induced base pair opening agrees with
   experimental observation on DNA base pair opening detected by potassium permanganate foot
   printing \cite{ref51}.  We will make numerical analysis of the discrete dynamical equation separately  to understand the discreteness effect which will be published elsewhere. It is also equally important to analyse the nonlinear
   dynamics of DNA double helix when the hydrogen bonding energy depends on the distance between the bases
   (inhomogeneity in hydrogen bonds) and the study is under progress. Also, the nonlinear dynamics
   study of open state configuration facilitated by enzymes (protein) is under progress.
  \section{Acknowledgements}
   The authors thank the anonymous referees for  useful comments.
 The major  portion of  the work is done within the framework of the Associateship Scheme of the Abdus Salam 
 International Centre  for Theoretical Physics, Trieste, Italy and the financial support is acknowledged. The work of M. D also  forms part of a major
  DST  project. V. V also thanks SBI for financial support.  
\appendix
\section{ Evaluation of  integrals in  Eq.(\ref{eq74}) using residue theorem}
In this appendix we evaluate the  integrals found in Eq.(\ref{eq74})
  using  standard residue theorem. Eq.(\ref{eq74}) is of the form

 \begin{eqnarray}
\Psi^{(1)} (\zeta,t_{0})&=&-\frac{1}{3\pi}\int_{-\infty}^{\infty}
\frac{dk}{(1+k^{2})^{3}} (1-k^{2}-2ik\tanh\zeta)
 e^{ik\zeta}\int_{-\infty}^{\infty}d\zeta(1-k^{2}\nonumber\\
&&+2ik\tanh\zeta)
 (\pi-6sech\zeta) sech\zeta\tanh\zeta\nonumber\\
&&\times \left[e^{-ik\zeta} -e^{i\frac{(1+k^{2})}{k}\alpha}e^{i\beta\zeta}\right].
\label{eqa1}
\end{eqnarray}
The evaluation of various integrals in Eq.~(\ref{eqa1}) can be
 facilitated by first finding the values of the
following two integrals \cite{ref52}.
\begin{eqnarray}
I_{1}=\int_{-\infty}^{\infty}  sech\zeta ~e^{i\chi\zeta}d\zeta, \label{eqa2}
\end{eqnarray}
\begin{eqnarray}
I_{2}=\int_{-\infty}^{\infty}  \tanh\zeta ~e^{i\chi\zeta}d\zeta, \label{eqa3}
\end{eqnarray}
where $\chi$ can take values $-k$ and $\beta$. 
 The integrand in $I_{1}$ is found to be analytic 
  everywhere except at the pole  $\zeta\rightarrow
i(2n+1)\frac{\pi}{2},~ n=0,1,2,....$  The residue of the  function
 $sech\zeta ~ e^{i\chi\zeta}$ can be written as
\begin{eqnarray}
Res(\zeta=i(2n+1)\frac{\pi}{2})= \overset{\infty}{\underset{n=0}{\sum}}
~\underset{\zeta\rightarrow i(2n+1)\frac{\pi}{2}}{lim} 
\frac{e^{i\chi\zeta}}{\frac{d}{d\zeta}(\cosh\zeta)} ,\label{eqa4}
\end{eqnarray}
which can be simplified to give 
\begin{eqnarray}
Res(\zeta=i(2n+1)\frac{\pi}{2})=\frac{1}{2i\cosh{\frac{\pi}{2}\chi}} .
\qquad\qquad\label{eqa5}
\end{eqnarray} 
 Thus the  integral value of $I_{1}$ in Eq.~(\ref{eqa2}) is written as
\begin{eqnarray}
I_{1}\equiv\int_{-\infty}^{\infty} sech\zeta e^{i\chi\zeta}d\zeta
=\frac{\pi}{\cosh{(\frac{\pi}{2}\chi)}}. \label{eqa6}
\end{eqnarray}
Similarly, using the same procedure we  evaluate the integral $I_{2}$ 
in which the integrand $\tanh\zeta ~ e^{i\chi \zeta}$ contains poles at $\zeta\rightarrow
i(2n+1)\frac{\pi}{2},~ n=0,1,2,...$ and obtain
\begin{eqnarray}
I_{2}\equiv\int_{-\infty}^{\infty} 
\tanh\zeta ~e^{i\chi\zeta} d\zeta=\frac{i\pi}{\sinh{(\frac{\pi}{2}\chi)}} .\label{eqa7}
\end{eqnarray}
Now using the value of the integral $I_{1}$  as given in Eq.~(\ref{eqa6}), we
 evaluate the following integrals  found in Eq.~(\ref{eqa1}) by 
integrating them  by parts successively to obtain 
\begin{subequations}
\begin{eqnarray}
\int_{-\infty}^{\infty} sech\zeta\tanh\zeta~ e^{i\chi\zeta}
d\zeta=\frac{i\pi \chi} {\cosh{(\frac{\pi}{2}\chi)}} ,\quad\quad\label{eqa8a}\\
\int_{-\infty}^{\infty}sech\zeta\tanh^{2}\zeta ~e^{i\chi\zeta} 
d\zeta=\frac{\pi(1-\chi^{2})}{2\cosh{(\frac{\pi}{2}\chi)}}.\qquad\label{eqa8b} 
\end{eqnarray}
\end{subequations}
Similarly, using Eq.~(\ref{eqa7}) we evaluate  the following 
integrals found in Eq.~(\ref{eqa1})  by making successive 
integrations by parts and obtain
\begin{subequations}
\begin{eqnarray}
\int_{-\infty}^{\infty} sech^{2}\zeta\tanh\zeta ~e^{i\chi\zeta}
d\zeta=\frac{i\pi \chi^{2}}{2\sinh{(\frac{\pi}{2}\chi)}} ,\qquad\label{eqa9a}\\
\int_{-\infty}^{\infty}  sech^{2}\zeta\tanh^{2}\zeta~ e^{i\chi\zeta}
d\zeta=\frac{\pi \chi(2-\chi^{2})}{6\sinh{(\frac{\pi}{2}\chi)}}.\quad\quad\label{eqa9b}
\end{eqnarray}
\end{subequations}
Here also $\chi$  can take values $-k$ and $\beta$. Substituting the values of the 
 integrals found in Eqs.~(\ref{eqa8a}), (\ref{eqa8b}) (\ref{eqa9a}) and 
(\ref{eqa9b}) in Eq.~(\ref{eqa1}), we obtain
\begin{eqnarray}
\Psi^{(1)} (\zeta,t_{0})&=&-\frac{i}{12}\left[\int_{-\infty}^{\infty}dk 
\frac{(1-k^{2}-2ik\tanh\zeta)}
{k^{2}(1+k^{2})^{2}}\left( \pi (1-v^{2}) k(1+k^{2})\right.\right.\nonumber\\
 && \times sech{\frac{\pi}{2}\beta} 
-\{(2-v)(1+v)^{2}(1+k^{2})^{2}-4 k^{2}(1+v+k^{2})\} \nonumber\\
&&\left. \times cosech{\frac{\pi}{2}\beta}\right)e^{i\frac{(1+k^{2})}{k}\alpha+ik\zeta}\nonumber\\ 
&&\left.+4\int_{-\infty}^{\infty}
dk \frac{(k^{2}-k^{4}-2ik^{3}\tanh\zeta)}{(1+k^{2})^{2}\sinh{\frac{\pi}{2}k}}e^{ik\zeta}
\right].\quad\label{eqa10}
\end{eqnarray}
\\

 Now, before writing down the final form of $\psi^{(1)}(\zeta,t_0)$ we evaluate the integrals
  with respect to $k$ in the right hand side of the above
 equation. For this first we rearrange 
 the integrand in
 Eq.(\ref{eqa10}) by simple multiplication and 
 call them   $I_{3} , I_{4}, I_{5}$ and $I_{6}$  as given below. 
 \begin{subequations}
 \begin{eqnarray}
 I_{3}&=&\int_{-\infty}^{\infty}
dk  \frac{(k^{2}-k^{4}-2 i k^{3}\tanh\zeta)}{{(1+k^{2})^{2}\sinh{\frac{\pi}{2}k}}}e^{ik\zeta},\label{eqa11a}\\
I_{4}&=&\int_{-\infty}^{\infty}dk 
\frac{(1-k^{2}-2ik\tanh\zeta) }
{(1+k^{2})^{2}\sinh{\frac{\pi}{2}\beta}}(1+v+k^{2})
 e^{i\frac{(1+k^{2})}{k}\alpha+ik\zeta},\label{eqa11b}\\
I_{5}&=&\int_{-\infty}^{\infty}dk 
\frac{(1-k^{2}-2ik\tanh\zeta)}{k^{2}\sinh{\frac{\pi}{2}\beta}} 
e^{i\frac{(1+k^{2})}{k}\alpha+ik\zeta},\qquad\label{eqa11c}\\ 
I_{6}&=&\int_{-\infty}^{\infty}dk 
\frac{(1-k^{2}-2ik\tanh\zeta)}
{k(1+k^{2})\cosh{\frac{\pi}{2}}\beta}
e^{i\frac{(1+k^{2})}{k}\alpha+ik\zeta}.\qquad\label{eqa11d}
 \end{eqnarray}
 \end{subequations}
 The integral $I_{3}$  can be evaluated by finding the residue  of the integrand
  $ \frac{(k^{2}-k^{4}-2ik^{3}\tanh\zeta)}{(1+k^{2})^{2}\sinh{k\frac{\pi}{2}}}e^{ik\zeta} $ at the poles
  $k=i$ of  order two
 and  at the simple pole   $k=2in$. The results are given by
 \begin{subequations}
 \begin{eqnarray}
 Res(k=i)=\frac{1}{2}(\zeta-2) sech\zeta,\qquad\qquad\qquad\qquad\qquad\qquad \qquad\quad\label{eqa12a}\\
 Res(k=2in)=\frac{40}{9\sqrt{2}~\pi} \sqrt{sech\zeta}e^{-\frac{3\zeta}{2}} 
 +\frac{32}{9\sqrt[3]{2}~\pi}  \tanh\zeta\sqrt[3]{sech\zeta}e^{-\frac{5\zeta}{3}}. \label{eqa12b}
 \end{eqnarray}
 \end{subequations}
 While writing the first term in  Eq.(\ref{eqa12b}), we have approximated  the results obtained
as a series in terms of suitable functions.
 Adding the above two residue values we obtain the value of the integral $I_{3}$ as
 \begin{eqnarray}
 I_{3}&=&2\pi i\left[\frac{1}{2}(\zeta-2) sech\zeta+\frac{40}{9\pi\sqrt{2}} 
 \sqrt{sech\zeta}e^{-\frac{3\zeta}{2}}+\frac{32}{9\sqrt[3]{2}~\pi}  \tanh\zeta\right.\nonumber\\
 &&\left.\times \sqrt[3]{sech\zeta}~ e^{-\frac{5\zeta}{3}}\right].
 \label{eqa13}
 \end{eqnarray}
  In the case of the integral $I_4$, the integrand possesses a second order pole at $k=i$ and there is
  one more simple pole at $k=\frac{(2n+1)}{(1-v)}(-i\pm\sqrt{\frac{1-v^{2}}{(2n+1)^{2}}-1})$
 which is however out of the contour. Thus the residue for the integrand function 
 $\frac{(1-k^{2}-2ik\tanh\zeta)}
{(1+k^{2})^{2}\sinh{\beta\frac{\pi}{2}}}(1+v+k^{2})e^{i\frac{(1+k^{2})}{k}\alpha+ik\zeta} $ is obtained as
\begin{eqnarray}
Res(k=i)=\frac{1}{2}(2+v\zeta+2v\alpha)sech\zeta,\label{eqa14}
\end{eqnarray}
and hence the right hand side of Eq.~(\ref{eqa14}) represents the value of the integral $I_{4}$.\\

The integrand of the integral $I_{5}$  namely
 $\frac{(1-k^{2}-2ik\tanh\zeta)}
{k^{2}\sinh{\beta\frac{\pi}{2}}}e^{i\frac{(1+k^{2})}{k}\alpha+ik\zeta} $  admits an essential singularity at
$k=0$, in addition to a simple pole at $k=\frac{(2n+1)}{(1-v)}(-i\pm\sqrt{\frac{1-v^{2}}{(2n+1)^{2}}-1})$
which is also out of the contour. The residue corresponding to the essential singularity is found by 
collecting the coefficient of $\frac{1}{k}$  after expanding the above function. Thus we obtain
\begin{eqnarray}
Res(k=0)&=&\overset{\infty}{\underset{n,n'=0}{\sum}} 
\frac{(-1)^{n+n'}(\frac{\pi}{2})^{2n-1} b^{n-1}(b-1)^{n}\alpha^{n'}}{(n'!)}
  (\zeta+\alpha)^{n'}\left[\frac{B_{n}}{(n!)^{2}(n'!)}\right.\nonumber\\
 && \times(2^{2n-1}-1)+\frac{(\frac{\pi}{2})^{2}b^{2}B_{n+1}}{(n+1)!^{2}(n'!)}
(2^{2n+1}-1)\left[1-2\tanh\zeta\right.\nonumber\\
&&\times\frac{(\zeta+\alpha)}{(n'+1)}
\left.+\frac{(\zeta+\alpha)^{2}}{(n'+1)(n'+2)}\right]
+\frac{(\frac{\pi}{2})^{4}b^{4}B_{n+2}}{(n+2)!(n+3)!}\nonumber\\
&&\times\frac{(n+1)(2^{2n+3}-1)}{(n'+2)!}\left[1-2\tanh\zeta\frac{(\zeta+\alpha)}{(n'+3)}\right.\nonumber\\
&&\left.\left.+\frac{(\zeta+\alpha)^{2}}{(n'+3)(n'+4)}\right]+...\right],
\label{eqa15}
\end{eqnarray}
where $B_{n}'$s are Bernoulli numbers and $b=\frac{(1+v)}{2}$. 
Thus the right hand side of Eq.~(\ref{eqa15}) gives the value of the integral $I_{5}$. In the case of  
 the integral $I_{6}$, the integrand
 $\frac{(1-k^{2}-2ik\tanh\zeta)}
{k(1+k^{2})\cosh{\frac{\pi}{2}}\beta}
e^{i\frac{(1+k^{2})}{k}\alpha+ik\zeta} $  possesses a
 second order pole at $k=i$  and an essential singularity at $k=0$. 
 In addition, we have a simple pole at 
 $k=\frac{(2n+1)}{(1-v)}(-i\pm\sqrt{\frac{1-v^{2}}{(2n+1)^{2}}-1})$
 which is again out of  the contour. The residues at the pole $k=i$ and at the 
 essential singularity $k=0$
 are respectively found to be
\begin{subequations}
\begin{eqnarray}
Res(k=i)&=&\frac{-1} {\pi v^{2}}(1-2 v\zeta-4v\alpha)
 sech\zeta,\label{eqa16a}\\
Res(k=0)&=&\overset{\infty}{\underset{m,n,n'=0}{\sum}}\overset{m}{\underset{j=0}{\sum}}
 \frac{(-1)^{n+n'}
(b-1)^{n}E_{n+m}(\frac{\pi}{2})^{2n+2m}b^{n+2m}\alpha^{n'} }{(n!)(n+2m)!(n')!(n'+2j)!}\nonumber\\
&&\times (\zeta+\alpha)^{n'+2j}\left[1-\frac{\frac{\pi}{2}^{2}b^{2}}{(n+1+2m)(n+2+2m)}\right.\nonumber\\
&&\left.-\frac{2(\frac{\pi}{2})^{2} b^{2}(\zeta+\alpha)\tanh\zeta}
{(n+1+2m)(n+2+2m)(n'+1+2j)}\right],\quad\label{eqa16b}
\end{eqnarray}
\end{subequations}
where $E_{n}'$s are Euler numbers. We evaluated the value of the residue given in (\ref{eqa16a}) using
Mathematica. 
The value of the integral $I_{6}$ is the sum of the right hand sides of Eqs.~(\ref{eqa16a}) and
~(\ref{eqa16b}).\\

Now the value of $\psi^{(1)}(\zeta,t_{0})$ is found by combining
 Eqs.~(\ref{eqa13}),~(\ref{eqa14}),~(\ref{eqa15}),
(\ref{eqa16a}) and ~(\ref{eqa16b}) and the final form is written as
\begin{eqnarray}
\Psi^{(1)}(\zeta,t_{0})&\approx&\frac{80}{27\sqrt{2}}\sqrt{sech\zeta}~e^{-\frac{3\zeta}{2}}
 +\frac{64}{27\sqrt[3]{2}}\tanh\zeta \sqrt[3]{sech\zeta}~e^{-\frac{5}{3}\zeta}\nonumber\\
 && +\frac{\pi}{6v^{2}}\left[2 v(1+v)\zeta+v^{2}+4\alpha v-1\right]sech\zeta. 
 \label{eqa17}
\end{eqnarray}
While writing the above, we have 
dropped few higher order terms  due to smallness in values 
that appeared in the  residue of the  essential singularity.
\section{ Evaluation of the integrals in Eq. (\ref{eq79}) using residue theorem}
In this appendix,  we evaluate the  integrals found in Eq.~(\ref{eq79}) using   standard
residue theorem as done in the previous  two cases. Eq.(\ref{eq79}) is written as
\begin{eqnarray}
\Psi^{(1)} (\zeta,t_{0})&\approx&\frac{1}{\pi}\int_{-\infty}^{\infty}
\frac{dk}{(1+k^{2})^{3}} (1-k^{2}-2ik\tanh\zeta) e^{ik\zeta} \int_{-\infty}^{\infty}
d\zeta(1-k^{2}\nonumber\\
&& +2ik\tanh\zeta)[\sin\zeta+(\cos\zeta -\frac{\pi^{2}}{8})\tanh\zeta]sech\zeta
 \nonumber\\
&&\times\{e^{i\frac{(1+k^{2})}{k}\alpha}e^{i\beta\zeta}-e^{-ik\zeta}\}
 +(1+v)sech\zeta\left[\int_{-\infty}^{\infty}
d\zeta ~\zeta^{2}\right.\nonumber\\
&& \times\left. [\sin\zeta+\cos\zeta]sech^{2}\zeta-\frac{\pi^{2}}{8}\int_{-\infty}^{\infty}
d\zeta ~\zeta^{2}sech^{2}\zeta\tanh\zeta \right] .\label{eqc1}
\end{eqnarray}
It may be verified that the last integral $\int_{-\infty}^{\infty}d\zeta~\zeta^{2} sech^{2}\zeta\tanh\zeta$
in the right hand side of Eq.(\ref{eqc1}) on evaluation vanishes. 
For evaluating  some of the  integrals found in Eq.~(\ref{eqc1}) we use the values of the
 integrals  given in
Eqs.~(\ref{eqa6}), (\ref{eqa8a}) and (\ref{eqa8b})  and also the value of the following integral.
\begin{eqnarray}
I_{8}=\int_{-\infty}^{\infty}  \zeta^{2} \tanh\zeta e^{i\zeta} d\zeta.\label{eqc2}
\end{eqnarray}
The integrand in Eq.(\ref{eqc2}) is found to be analytic everywhere except at the pole
$\zeta=i(2n+1)\frac{\pi}{2}$, n=0,1,2,... The residue of the function
 $ \zeta^{2} \tanh\zeta e^{i\zeta}$
is  then found to be $
-\frac{\pi^{2}}{8\sinh(\frac{\pi}{2})}
\left[1+2\frac{e^{-\frac{\pi}{2}}}{\sinh(\frac{\pi}{2})}
+\frac{e^{\frac{\pi}{2}}(1-e^{-2\pi\chi})}{2\sinh^{3}(\frac{\pi}{2})}\right]$ and hence the value of the
integral $I_8$ is written as 
\begin{eqnarray}
I_{8}\equiv\int_{-\infty}^{\infty} \zeta^{2} \tanh\zeta e^{i\zeta}
d\zeta=-\frac{i\pi^{3}}{4\sinh(\frac{\pi}{2})}
\left[1+2\frac{e^{-\frac{\pi}{2}}}{\sinh(\frac{\pi}{2})}
+\frac{e^{\frac{\pi}{2}}(1-e^{-2\pi})}{2\sinh^{3}(\frac{\pi}{2})}\right].\label{eqc3}
\end{eqnarray}
 Now, using the value of the integral given in Eq.~(\ref{eqc3}), we evaluate the following integrals found in
 Eq.~(\ref{eqc1}) by integrating them by parts successively.
 \begin{subequations}
  \begin{eqnarray}
 \int_{-\infty}^{\infty}d\zeta\zeta^{2}sech^{2}\zeta e^{\pm
 i\zeta}&=&\frac{\pi^{2}}{\sinh\frac{\pi}{2}}\left[\pm(1\mp\frac{\pi}{4})
 +\frac{(1\mp\frac{\pi}{2})e^{\mp\frac{\pi}{2}}}{\sinh\frac{\pi}{2}}\right.\nonumber\\
  &&\left.\pm\frac{\pi e^{\pm\frac{\pi}{2}}(1-e^{\mp2\pi})}
{\sinh^{3}\frac{\pi}{2}}\right],\label{eqc4a}\\
\int_{-\infty}^{\infty}d\zeta\zeta^{2}sech^{2}\zeta\tanh\zeta e^{ \pm i\zeta}&=&\frac{i\pi}{\sinh\frac{\pi}{2}}
\left[(\pi\mp1\mp\frac{\pi^{2}}{8})\pm
\frac{\pi(2\mp\frac{\pi}{2})
e^{\mp\frac{\pi}{2}}}{2\sinh\frac{\pi}{2}}\right.\nonumber\\
&&\left.-\frac{i\pi^{2}e^{\pm\frac{\pi}{2}}(1-e^{\mp 2\pi})}{16\sinh^{3}\frac{\pi}{2}}\right].\quad\label{eqc4b}
\end{eqnarray}
  \end{subequations}  
  On substituting Eqs.~(\ref{eqc4a}) and (\ref{eqc4b}) in Eq.~(\ref{eqc1}) we obtain
  \begin{eqnarray} 
\Psi^{(1)}(\zeta,t_{0})&=&\frac{(1+v)i\pi^{3}}{16\sinh\frac{\pi}{2} }
\left[\frac{\cosh\frac{3\pi}{2}}{\sinh^{3}\frac{\pi}{2}}
 -\cosh\frac{\pi}{2}\left(\frac{4}{\sinh\frac{\pi}{2}}+1\right )
\right] sech\zeta\nonumber\\
  &&+\frac{i}{2}\int_{-\infty}^{\infty}dk
  \frac{(1-k^{2}-2ik\tanh\zeta)}{k(1+k^{2})^{3}}
  e^{\frac{i(1+k^{2})\alpha}{k}+ik\zeta}\left[ \{k^{2}+b(1-b)\right.\nonumber\\
&&\times(1+k^{2})^{2}\}
  \left(sech{\frac{\pi}{2}(\beta+1)}
 +sech{\frac{\pi}{2}(\beta-1)}\right)
 -\frac{\pi^{2}}{4}b(1-b)\nonumber\\
&&\left.\times(1+k^{2})^{2}sech{\frac{\pi}{2}\beta}\right]
-\frac{i}{2}\int_{-\infty}^{\infty}
dk \frac{ (k-k^{3}-2ik\tanh\zeta)}{(1+k^{2})^{3}}\nonumber\\
 &&\times e^{ik\zeta}\left(sech{\frac{\pi}{2}(1-k)}+sech{\frac{\pi}{2}(1+k)}\right).
 \label{eqc5}
  \end{eqnarray}
  Before evaluating the  integrals  in Eq.(\ref{eqc5}) we rewrite the same appropriately.
   We then evaluate the integrals  in Eq.~(\ref{eqc5}) one by one by finding the values of the residues
     at the pole $(k=i)$ at different orders. 
    The residue for the integrand  function $\left(sech{\frac{\pi}{2}(\beta+1)}
 +sech{\frac{\pi}{2}(\beta-1)}\right)\frac{k(1-k^{2}-2ik\tanh\zeta)}
 {(1+k^{2})^{3}}e^{\frac{i(1+k^{2})\alpha}{k}+ik\zeta}$ at the pole $ k=i$ of order three is found to be
 \begin{eqnarray}
Res(k=i)=-\frac{\pi}{16} [2(2b-1)\zeta+4(2b-1)\alpha+1)]
sech\zeta.\label{eqc6}
 \end{eqnarray} 
 Next, we find the residue for the function 
 $e^{\frac{i(1+k^{2})\alpha}{k}+ik\zeta}\frac{(1-k^{2}-2ik\tanh\zeta)}
{k(1+k^{2})\cosh\beta\frac{\pi}{2}}$ at the  simple poles
 $k=i$ and   $k=\frac{(2n+1)}{(1-v)}(-i\pm\sqrt{\frac{1-v^{2}}{(2n+1)^{2}}-1})$ ( which is out of
contour). Further, at $ k=0$ there is an essential singularity, the residue of which is not shown here due its
unwieldy form.  Thus
 the residue for the above function  at the first order pole $ k=i$  is given by
\begin{eqnarray}
Res(k=i)=\frac{1}{\pi (1-2b)^{2}} [2(2b-1)\zeta 
+4(2b-1)\alpha-1] sech\zeta.\label{eqc7}
\end{eqnarray}
 The residues for the function $\left(sech{\frac{\pi}{2}(k-1)}
 +sech{\frac{\pi}{2}(k+1)}\right)\frac{(k-k^{3}-2ik^{2}\tanh\zeta)}{(1+k^{2})^{3}}e^{ik\zeta}
$ at the 
pole  $k=i$ of order  three and at the simple poles  $k=i(2n+1)+1$ and $k=i(2n+1)-1$
are written as 
\begin{subequations} 
\begin{eqnarray}
Res(k=i)&=&\frac{\pi}{16}(2\zeta-1)sech\zeta, \label{eqc8a}\\
Res[k=i(2n+1)+1]&=&\frac{-2 i}
{\pi}e^{(i-1)\zeta}\sum_{n=0}
\frac{(-1)^{-n}[1+i(2n+1)]e^{-2n\zeta}}{[2-(2n+1)^{2}+2i(2n+1)]^{3}}\nonumber\\
&&\times\{[(2n+1)^{2}-2 i(2n+1)]\nonumber\\
&&-2i\tanh\zeta [1+i(2n+1)]\},\label{eqc8b}\\
Res[k=i(2n+1)-1]&=&\frac{2 i}
{\pi}e^{-(1+i)\zeta}\sum_{n=0}
\frac{(-1)^{-n}[-1+i(2n+1)]e^{-2n\zeta}}{[(2n+1)^{2}-2+2i(2n+1)]^{3}}\nonumber\\
&&\times\{[(2n+1)^{2}+2 i(2n+1)]\nonumber\\
&&-2i\tanh\zeta [-1+i(2n+1)]\}.\label{eqc8c}
\end{eqnarray}
\end{subequations}
 It can be verified  that the residue of the function $\left(sech{\frac{\pi}{2}(\beta+1)}
 +sech{\frac{\pi}{2}(\beta-1)}\right)$ $\frac{(1-k^{2}-2ik\tanh\zeta)}{k(1+k^{2})}
 e^{\frac{i(1+k^{2})\alpha}{k}+ik\zeta}
 $ at $k=i$ vanishes.  On summing up the values  of the residues given in Eqs. (\ref{eqc6}), (\ref{eqc7})
 and (B8) (dropping higher order terms in Eqs. (\ref{eqc8b}) and (\ref{eqc8c}) while
 adding) we obtain  the following expression for $\psi ^{(1)}(\zeta,t_{0})$.
 \begin{eqnarray}
 \Psi^{(1)}(\zeta,t_{0})&\approx&\frac{\pi^{2}}{4(1-2b)^{2}}sech\zeta
 \left[b(2b-1)\zeta
 +\{b(2\alpha+b-1)-\alpha\}+\frac{4}{122825}\right.\nonumber\\
  &&\times\{ (\cosh 2\zeta-\sinh 2\zeta)
 \left[25\cosh 2\zeta(523\cos\zeta-1512\sin\zeta)+289\right.\nonumber\\
 &&(19\cos\zeta+8\sin\zeta)+15(766cos\zeta
\left.\left. -2393\sin\zeta)\sinh 2\zeta\right]\}\right] .\label{eqc9}
 \end{eqnarray}
  As  residues due to the singularities in the plane except along the
 imaginary axis, lead to secular terms in the solutions, we take into
 account only  singularities along the imaginary axis  that is at $k=i$, dropping 
 residues due to the other two singularities  at  $k=i(2n+1)+1$ and
 $k=i(2n+1)-1$. Thus the  first order correction $\psi^{(1)}(\zeta,t_{0})$ given in Eq.(\ref{eqc9}) finally
  becomes
 \begin{eqnarray}
 \Psi^{(1)}(\zeta,t_{0})&\approx&\frac{\pi^{2}}{4(1-2b)^{2}}
 \left[b(2b-1)\zeta
 +\{b(2\alpha+b-1)-\alpha\}\right]sech\zeta.
 \end{eqnarray} 

\end{document}